\documentclass[twocolumn]{autart}

\usepackage{amsfonts}
\usepackage{physics}
\usepackage{enumerate}
\usepackage{xcolor}
\usepackage{balance}
\usepackage{tabularx}
\usepackage{natbib}        
\def\RR{\mathbb{R}}

\def\EE{\mathbb{E}}
\def\PP{\mathbb{P}}
\def\I{\mathcal{I}}

\def\U{\mathcal{U}}
\def\xhat{\hat{x}}

\def\bs{\boldsymbol}
\def\bxhat{\hat{\bf x}}
\def\bx{{\bf x}}
\def\bu{{\bf u}}
\def\bv{{\bf v}}
\def\bw{{\bf w}}
\def\Q{{\bf Q}}
\def\R{{\bf R}}
\def\ones{{\bf 1}}
\def\defeq{:=}
\def\tail{\circ}

\def\bpf{\textnormal{\textit{Proof:}\hspace{1ex}}}
\def\epf{\hfill \mbox{\qed}}

\DeclareMathOperator{\trloc}{tr}
\DeclareMathOperator{\diag}{diag}
\DeclareMathOperator{\vvec}{vec}
\DeclareMathOperator{\reshape}{mat}

\allowdisplaybreaks[4]

\newtheorem{assumption}{Assumption}

\makeatletter
\g@addto@macro\normalsize{%
  \setlength\abovedisplayskip{2.0pt}
  \setlength\belowdisplayskip{2.0pt}
  \setlength\abovedisplayshortskip{1.5pt}
  \setlength\belowdisplayshortskip{1.5pt}
}
\makeatother

\begin{document}
\begin{frontmatter}

\title{Stochastic output feedback MPC with intermittent observations\thanksref{footnoteinfo}}

\author[shuhao]{Shuhao Yan\corauthref{cor}}\ead{sy499@cornell.edu}
\quad
\author[Oxford]{Mark Cannon}\ead{mark.cannon@eng.ox.ac.uk}
\quad
\author[Oxford]{Paul J. Goulart}\ead{paul.goulart@eng.ox.ac.uk}

\address[shuhao]{School of Electrical and Computer Engineering, Cornell University, Ithaca, NY, 14853, USA}
\address[Oxford]{Department of Engineering Science, University of Oxford, Oxford OX1 3PJ, UK}

\corauth[cor]{Corresponding author.}
\thanks[footnoteinfo]{A preliminary version of this paper was presented at the 21st IFAC World Congress, July 12-17, 2020.}

\begin{abstract}                
This paper designs a model predictive control (MPC) law for constrained linear systems with stochastic additive disturbances and noisy measurements, minimising a discounted cost subject to a discounted expectation constraint.
It is assumed that sensor data is lost with a known probability. Taking into account the data losses modelled by a Bernoulli process, we parameterise the predicted control policy as an affine function of future observations and obtain a convex linear-quadratic optimal control problem.
Constraint satisfaction and a discounted cost bound are ensured without imposing bounds on the distributions of the disturbance and noise inputs.
In addition, the average long-run undiscounted closed loop cost is shown to be finite if the discount factor takes appropriate values. 
We analyse robustness of the proposed control law with respect to possible uncertainties in the arrival probability of sensor data and we bound the impact of these uncertainties on constraint satisfaction and the discounted cost.
Numerical simulations are provided to illustrate these results.
\end{abstract}


\end{frontmatter}

\section{Introduction}\label{sec:intro}
Applications of wireless sensor networks are becoming omnipresent, for example in healthcare monitoring, weather forecasting, autonomous driving and building management systems. These networks can be time-varying and subject to congestion.
Networked control systems that rely on sensor signals transmitted over communication channels must tolerate communication delays and data losses. Such measurement uncertainties pose additional challenges for estimator and controller design, especially when constraints are present.

A large body of work exists on control problems involving estimation with intermittent observations.
This is largely concerned with the Random Riccati Equation \citep{Wang:99}, and the main issue addressed is boundedness of the state estimation error covariance.
In \cite{sinopoli:04}, packet losses are modelled by an independent and identically distributed (i.i.d.) Bernoulli process and fundamental results are reported, such as the existence of and bounds on the critical value for the arrival probability of the observation update and convergence properties of the algebraic Riccati equation for Kalman filters with intermittent observations.
This is the basis of further work in \cite{kar:12} showing that the estimation error covariance sequence $\Sigma_0,\Sigma_1,\ldots$ converges in distribution to a unique invariant distribution with finite mean $\bar{\Sigma}$ when the observation arrival probability exceeds the critical value.
In \cite{shi:10}, the boundedness of the estimation error covariance is studied from
a probabilistic perspective and bounds on the probability of $\Sigma_k \preceq \bar{\Sigma}$ (henceforth denoted $\PP\{\Sigma_k \preceq \bar{\Sigma} \}$) are derived.
A similar metric is used in \cite{mo:12} to study the boundedness of estimation error covariance.
This work considers the tail distribution of the trace of the estimation error covariance and the decay rate of $\PP\{\trloc(\Sigma_k) >\trloc(\bar\Sigma)\}$ (where $\trloc(\cdot)$ denotes the trace of a matrix) with packet losses being modelled by a Markov process. For the case of Markovian packet losses, it is possible to account for temporal correlation of communication channel conditions,
but the stability analysis is more involved without an i.i.d. assumption \citep{you:2011,huang:2007}.
In contrast to the aforementioned work focusing on limiting behaviours of estimation error covariance, \cite{huang2015escape} provide a different perspective by analysing transient properties. This work considers an intermittent quantised communication link and characterises the cumulative probability distribution function of the escape time at which system states or state estimates first exit a given compact set.

The results mentioned above do not easily carry over to constrained cases, and 
problems involving constrained estimation and control under intermittent and noisy observations have received less attention.
In \cite{ren:2013}, the problem of minimising an average estimation error covariance over a finite horizon is studied subject to an energy constraint on the remote sensor. \cite{leong:2012} consider a similar problem subject to an average transmission power constraint, while the probability of packet losses at each time instant is conditional on transmission power. In \cite{mishra:2020}, the problem of controlling stable linear systems with Gaussian disturbances and measurement noise is considered subject to hard input constraints, where both sensor and control channels are unreliable. This work assumes that remote sensors equipped with computing capabilities send state estimates instead of raw sensor measurements, which could be costly and unreasonable in certain situations.

Assuming sensor measurements will be lost with a known probability according to a Bernoulli process, this paper proposes an MPC algorithm with state estimation
to minimise a discounted cost subject to a discounted expectation constraint. The system model has additive disturbances and measurement noise with probability distributions that are not assumed to be known exactly and may have infinite support.
The discount setting is common to many control problems \citep[e.g.][]{bertsekas:95,blackwell1965discounted,VanParys:13,KOUVARITAKIS:03,KAMGARPOUR:17,feinberg:1999}, and an appropriate discount factor can provide stability guarantees \citep{postoyan:17}.
In financial engineering applications and economics, discounting is typically used to cost future activities at the present time. 
More generally, the effects of uncertain predictions can be mitigated by discounting controller objectives and constraints, reflecting the considerations that near-term benefits are more important than potential future gains and that current actions should not imperil possible future development \citep{kouvaritakis2006TAC}. Examples include sustainable development \citep{kouvaritakis2006TAC}, public policy analysis \citep{dasgupta08discounting} and ecosystem management \citep{clark1973profit} where different discount factors can affect harvesting policies and species populations.
Relative to our earlier work in~\cite{yan:20}, this paper includes additional results on 
the terminal term in the cost, on stability in the discount setting and on robustness analysis with respect to uncertainties in the arrival probability of sensor data, plus more extensive numerical studies.
The main contributions of this paper are summarised as follows:
\begin{itemize}
\item 
The discount factors ensure both the cost function and constraint are well defined without bounds on the disturbance and measurement noise, and achieve a trade-off between short-term and long-term performance;
\item 
We parameterise predicted control inputs as affine functions of future output measurements and show that the problem of optimising the associated feedback gains is convex. This allows the distributions of future states to be controlled explicitly even when output measurements are lost;
\item 
We derive a bound on the discounted cost and ensure constraint satisfaction for the closed loop system;
\item 
We provide two different conditions for ensuring stochastic stability in closed loop;
\item We investigate the impact of uncertainty in the arrival probability of sensor data on constraint satisfaction and on the discounted cost.
\end{itemize}

The paper is organised as follows. The control problem is described in Section \ref{sec:problem description}, and the controller parameterisation and implementation are introduced in Section \ref{sec:controller}.
Section \ref{sec:MPC_predictions} derives the expressions of
predicted state and control sequences via their first and second moments. Using these expressions, we formulate the MPC optimisation problem 
in Section \ref{sec:cost and constraints}.
Section \ref{sec:properties}
derives a closed loop discounted cost bound and analyses constraint satisfaction. The averaged undiscounted closed loop cost accumulated over an infinite horizon is shown to be finite and an explicit bound is given in Section~\ref{sec:strengthened stability}. 
We conduct a robustness analysis with respect to possible uncertainties in the arrival probability of sensor measurements in Section \ref{sec:robustness analysis}.
Sections~\ref{sec:example} and~\ref{sec:conclusion} provide a numerical example and conclusions.

\textit{Notation}:
The $n\times n$ identity matrix is $I_{n\times n}$, and the $n\times m$ matrix with all elements equal to $1$ is $\ones_{n\times m}$.
%
The vectorised form of a matrix $A= [a_1 \ \cdots \ a_n]$ is $\vvec(A)\defeq[a_1^\top \ \cdots \ a_n^\top]^\top$ and $A\otimes B$ is the Kronecker product.
The Moore–Penrose inverse of a matrix $A$ is denoted as $A^\dagger$.
The Euclidean norm is $\| x \|$ and, for a matrix $Q$, $Q\succ 0$ ($Q\succeq 0$) indicates that $Q$ is positive definite (semidefinite) and $\|x\|_Q^2 \defeq x^\top Q x$. The notation $\lim_{x\to a^-} f(x)$ denotes the one-sided limit of $f(x)$ as $x$ approaches $a$ from the left and
$\reshape(x,[r,c])$ denotes the columnwise operation to reshape a vector $x$ of appropriate length into an $r$-by-$c$ matrix.

\section{Problem description}\label{sec:problem description}

\subsection{System model and feedback information} \label{section:system model}

We assume a system with linear discrete time dynamics
\begin{subequations}\label{eq:system}
\begin{align}
x_{k+1} &= A x_k + B u_k + D w_k,\\
y_k &= Cx_k + v_k,\\
z_k &= \gamma_k y_k,
\end{align}
\end{subequations}
where $x_k\in\RR^{n_x}$, $u_k\in\RR^{n_u}$, $y_k\in\RR^{n_y}$, $z_k\in\RR^{n_y}$, $w_k\in\RR^{n_w}$, $v_k\in\RR^{n_y}$ are the state, control input, sensor measurement, measurement information received by the controller, disturbance and the measurement noise, respectively.
The disturbance, measurement noise and packet loss sequences, $\{w_k\}_{k=0}^\infty$, $\{v_k\}_{k=0}^\infty$ and $\{\gamma_k\}_{k=0}^\infty$, are assumed to be i.i.d.\ with
\begin{alignat*}{2}
& \EE\{w_k\} = 0 , & \qquad & \EE\{w_kw_k^\top\} =: \Sigma_w \succeq 0 , \\
& \EE\{v_k\} = 0 , & & \EE\{v_kv_k^\top\} =: \Sigma_v \succ 0 ,\\
& \PP\{\gamma_k= 0\} = 1-\lambda ,  & & \PP\{\gamma_k = 1\} =:  \lambda .
\end{alignat*}
The variable $\gamma_k\in\{0,1\}$ indicates whether sensor data at the $k$th sampling instant is received by the controller.
The information assumed available to the controller at the $k$th time instant consists of
$\{u_i\}_{i=0}^{k-1}$, $\{(z_i,\gamma_i)\}_{i=0}^k$,
the initial mean $\EE\{x_0\}=:\xhat_0$, and model state covariance $\EE\{(x_0-\xhat_0)(x_0 - \xhat_0)^\top\} =: \Sigma_0$.

We define the information sets
\[
\I_{k} \defeq \{\I_{k-1}, (z_k,\gamma_k)\},
\quad
\U_{k} \defeq  \{\U_{k-1},u_k\},
\]
for all $k\geq 0$, where $\I_{-1} \defeq  \{\xhat_0,\Sigma_0\}$, $\U_{-1} \defeq \{\,\}$.  Finally, we define conditional expectation operators as
\[
\EE_k\{ \cdot \} \defeq \EE\{ \cdot \, | \, \U_{k-1},\I_{k-1}\},
\quad
\EE\{ \cdot \} \defeq \EE_0\{\cdot\}.
\]

\begin{assumption}\label{assumption:ctrb_and_obsv}
The pair $(A,B)$ is stabilisable, and $(A,C)$ is detectable.
\end{assumption}

\subsection{Optimal control problem}
We consider the problem of minimising the discounted sum of expected future values of ${\|x_{k}\|_Q^2 + \|u_{k}\|_R^2}$, where $Q\succeq 0$ and $R\succ 0$. The minimisation is performed subject to a constraint on the discounted sum of second moments of an auxiliary output, defined for a given matrix $H$  by $\xi_k = H x_{k}$, so the optimal control problem is
\begin{align}
\min \
& \sum_{k=0}^{\infty} \beta_1^k \EE\{  \|x_{k}\|_Q^2 + \|u_{k}\|_R^2 \}
\nonumber \\
\text{s.t.} \
& \sum_{k=0}^\infty \beta_2^k \EE\{ \| H x_{k}\|^2 \} \leq \epsilon .
\label{eq:orig_opt_problem}
\end{align}
Here $\beta_1\in(0,1)$ and $\beta_2\in(0,1)$ are discount factors and $\epsilon$ is a given constraint threshold.
For simplicity (with the understanding that the general case of $\beta_1 \neq \beta_2$ can be treated similarly),
we set $\beta_1=\beta_2=\beta$ for the remainder of the paper with the exception of Sections~\ref{sec:cl_bound_via_limits} and~\ref{sec:example}.
\begin{rem}\label{remark:special feature discounting}
The constraint in \eqref{eq:orig_opt_problem} can be interpreted as an energy constraint, for example, on signals transmitted over a network, and we note that constraints involving first order moments can be handled with minor modifications to our approach.
The discount factor in \eqref{eq:orig_opt_problem} weights performance closer to the initial time more
heavily than future behaviour. In contrast to average constraints over an infinite horizon \citep[e.g.][]{leong:2012}, 
this discounted constraint allows for a trade-off between short-term and long-term
behaviours and 
indicates preferential allocation of the energy budget towards time instants nearer the initial time.
\end{rem}

We will employ a receding horizon control law based on a finite-horizon control policy in the form
\begin{equation}
u_{i|k} = \kappa_i(\theta_k ,\U_{k+i-1},\I_{k+i}) \label{eq:kappa function}
\end{equation}
where $u_{i|k}$ for $i=0,1,\ldots$ is the prediction of $u_{k+i}$ at time instant $k$, and $\theta_k$ is a vector of controller parameters at time instant $k$. The dependence of $\kappa_i(\cdot)$ on the sets $\U_{k+i-1}$ and $\I_{k+i}$ ensures causality and the dependence on $\theta_k$ is chosen so that the optimal parameter vector, denoted $\theta_k^\ast$, will be the solution of a convex problem.

\begin{assumption}\label{assumption:info}
(i) The probability, $\lambda$, of successfully receiving sensor measurements is known and time-invariant. (ii) When $\theta_k^\ast$ is computed, $(z_{k+i},\gamma_{k+i})$ are unknown for all $i\geq 0$.
\end{assumption}


Thus $\theta_k^\ast$ is computed online prior to the $k$th sampling instant using knowledge of $\U_{k-1}$ and $\I_{k-1}$, while $(z_k, \gamma_k)$ is known when the control law is implemented, via
\[
u_k = \kappa_0(\theta_k^\ast , \U_{k-1}, \I_{k}) .
\]
In order to address \eqref{eq:orig_opt_problem}, we define an MPC optimisation problem to be solved at time instant $k$ as follows
\begin{equation}\label{eq:opt}
\begin{aligned}
\theta_k^\ast = \arg\min_{\theta_k} \
& \sum_{i=0}^{\infty} \beta^i \EE_k\{  \|x_{i|k}\|_Q^2 + \|u_{i|k}\|_R^2 \}  \\
\text{s.t.} \
& \sum_{i=0}^\infty \beta^i \EE_k\{ \| H x_{i|k} \|^2\} \leq \mu_{k} .
\end{aligned}
\end{equation}
Here $\mu_0=\epsilon$, and $\mu_k$ for $k>0$ is chosen as described in Section~\ref{sec:properties} to ensure that (\ref{eq:opt}) is recursively feasible and that the closed loop system satisfies the constraint in~(\ref{eq:orig_opt_problem}).

\section{Controller design} \label{sec:controller}
\subsection{Controller parameterisation}\label{sec:control_parameterisation}
Consider the output feedback control law defined by an observer and an affine feedback law:
\begin{subequations}\label{eq:basic_control_law}
\begin{align}
\xhat_k &= A \tilde{x}_{k-1} + Bu_{k-1}, \\
\tilde{x}_k &= \xhat_{k} + \gamma_k M (y_k - C\xhat_k), \\
u_k &= K \tilde{x}_k  + c_k
\end{align}
\end{subequations}
with $\xhat_0 \!=\! \EE\{x_0\}$, where $\xhat_k$ and $\tilde{x}_k$ are the prior estimate and the posterior estimate of $x_k$, 
defined as $\xhat_k := \EE_k\{x_k\}$ and $\tilde{x}_k := \EE_{k+1}\{x_k\}$. Equations (\ref{eq:basic_control_law}a) and (\ref{eq:basic_control_law}b) are referred to as a prediction step and an update step respectively. Together with the assumption on $w_k$ and $v_k$, these two steps ensure the estimator is unbiased, namely, $\EE\{x_k\}=\EE\{\xhat_k\}=\EE\{\tilde{x}_k\}$.

If we fix the gains $M$ and $K$ and choose the decision variables in problem (\ref{eq:opt}) as $\theta_k = \{c_{0|k},\ldots,c_{N-1|k}\}$ for a fixed $N$, with the predicted control sequence defined as $u_{i|k} = K \tilde{x}_{i|k}  + c_{i|k}$, we would obtain a simplistic parameterisation of the predicted control law $\kappa_i(\cdot)$ in~\eqref{eq:kappa function}.
Although in this case, the number of decision variables increases only linearly with $N$, the parameters $\{c_{0|k},\ldots,c_{N-1|k}\}$ constitute an open loop control sequence that does not vary with the realisations of future measurement noise and disturbances. This is likely to yield poor performance and small feasible sets of initial conditions when the probability of packet loss is non-zero.

Instead we use a parameterisation that allows optimisation of the dependence of predicted control inputs on future realisations of model uncertainty. This allows the predicted probability distributions of states and control inputs to be controlled explicitly, thus providing flexibility to balance conflicting requirements for performance and constraint satisfaction.
Analogously to the case of predicted control laws in which state feedback gains are optimisation variables~\citep{lofberg:03,goulart:06}, the cost and constraints of problem~(\ref{eq:opt}) become nonconvex if time-varying gains $M$, $K$ are chosen as decision variables.
However, if the predicted control sequence is parameterised in terms of affine functions of the future output measurements received by the controller \citep{BBN:2006}, then the first and second moments of predicted states and inputs are convex functions of controller parameters.
By allowing arbitrary linear dependence of $\kappa_i(\cdot)$ on the received sensor measurements,
this approach makes it possible to optimise the future control sequence at every sampling instant, including those at which information from sensors is lost.

Therefore, let the $i$-step-ahead predicted control input $u_{i|k}$ be defined for all $i = 0,1,\ldots$ as
\begin{subequations}\label{eq:predicted_control_law}
\begin{align}
u_{i|k} \!&= \!K \xhat_{i|k} + c_{i|k} + d_{i|k},\\
d_{i|k} \!&= \!\gamma_{0|k}L_{i,0|k} (y_{0|k} \!-\! C\xhat_{0|k})  \!+\! \gamma_{1|k}L_{i,1|k} (y_{1|k} \!-\! C\xhat_{1|k})  \nonumber \\
&\quad + \cdots +
\gamma_{i|k}L_{i,i|k} (y_{i|k} - C\xhat_{i|k}),
\\
\xhat_{i+1|k} \!&=\! A\xhat_{i|k} + B u_{i|k} + \gamma_{i|k} AM (y_{i|k} - C\xhat_{i|k} )
\end{align}
\end{subequations}
where $c_{i|k} = 0$ and $L_{i,j|k} = 0$ $\forall i\geq N$. Here $\gamma_{i|k}$ and $y_{i|k}$ are random variables, denoting the $i$-step-ahead predicted packet loss and sensor measurement at time instant $k$, respectively, and therefore the probability distribution of $\gamma_{i|k}$ is chosen to be the same as that of $\gamma_k$.
Then, $\forall i = 0,1,\ldots$, the predicted state estimate satisfies
\begin{equation}\label{eq:nominal_state_estimate}
\xhat_{i+1|k} = \Phi \xhat_{i|k} \!+\! B(c_{i|k} \!+\! d_{i|k}) \!+ \gamma_{i|k} A M (y_{i|k} \!-\! C \xhat_{i|k})
\end{equation}
with $\Phi \defeq A+BK$. Since $x_{i+1|k} = A x_{i|k} + Bu_{i|k} + Dw_{i|k}$, the predicted estimation error evolves according to
\begin{equation}\label{eq:state_prediction}
x_{i+1|k} \!- \xhat_{i+1|k} \!\!=\!\! \Psi_{i|k}(x_{i|k} \!- \xhat_{i|k})
- \gamma_{i|k}AMv_{i|k} \!+\! D w_{i|k}
\end{equation}
where $\Psi_{i|k} \defeq A (I - \gamma_{i|k}M C)$.
These relationships allow the first and second moments of $x_{i|k}$ to be determined in terms of the decision variable $\theta_k$, which consists of the parameters
$\{c_{0|k},\ldots,c_{N-1|k}\}$
and feedback gains
$L_{0,0|k}$, $\{L_{1,0|k},L_{1,1|k}\}$, $\ldots, \{L_{N-1,0|k},\ldots,L_{N-1,N-1|k}\}$.

The gains $K$, $M$ in (\ref{eq:predicted_control_law}a-c) are fixed and determined of\mbox{}f\mbox{}line, satisfying the following assumption.

\begin{assumption}\label{assumption:stabilising_gains}
Gains $K$ and $M$ are chosen so that
$\xi_{i+1} = (A+BK) \xi_i$ is asymptotically stable and $\xi_{i+1} = A(I - \gamma_iMC) \xi_i$ is mean-square stable (MSS) \citep{Kushner:71}.
\end{assumption}

\begin{rem}\label{remark:choices of K and M}
Matrices $K$ and $M$ exist satisfying Assumption~\ref{assumption:stabilising_gains} if Assumption~\ref{assumption:ctrb_and_obsv} holds and if the  probability, $\lambda$, of successfully receiving a sensor measurement is greater than a critical value \citep[e.g.][]{sinopoli:04}. Suitable choices for $K$, $M$
are the optimal gains for (\ref{eq:opt}) in the absence of constraints, or the certainty equivalent LQ feedback gain for a problem with state and control weighting matrices $Q$ and $R$ and the steady state Kalman filter gain~\citep{sinopoli:04}. Note that time-varying gains $K_k$, $M_k$ can be used within the framework of this paper, provided the dependence on $\gamma_k$ is known in advance.
\end{rem}

\subsection{Controller implementation}
\label{sec:control_implementation}

The control law is implemented as follows:
\begin{enumerate}[(i)]
\item
Given $\mathcal{U}_{k-1}$ and $\I_{k-1}$, solve problem (\ref{eq:opt}) for $\theta^\ast_k$;
\item
Given $\gamma_k$ and $z_k=\gamma_k y_k$:
\begin{enumerate}[(a)]
\item
apply the control input
\begin{equation} \label{eq:mpc controller}
u_k = K\xhat_k + c_{0|k}^\ast + \gamma_k L^\ast_{0,0|k}(y_k - C\xhat_k),
\end{equation}
\item
update the state estimate
\begin{equation}
\xhat_{k+1} = A\xhat_k + B u_k + \gamma_k A M(y_k - C\xhat_k). \label{eq:cl loop priori estimate update}
\end{equation}
\end{enumerate}
\end{enumerate}
Note that this receding horizon control law includes (\ref{eq:basic_control_law}) as a special case, since $u_{k}$ and $\xhat_{k+1}$ in step (ii) are equal to their counterparts in (\ref{eq:basic_control_law}) if $(c_{0|k}^\ast,L_{0,0|k}^\ast) = (c_k,KM)$.

\section{Predicted state and control sequences}
\label{sec:MPC_predictions}
To simplify notation, we define vectorised sequences, with
$\bx_k$ denoting the vectorised true state sequence $\{x_{i|k}\}_{i=0}^{N-1}$,
$\bxhat_k$ the estimated state sequence $\{\xhat_{i|k}\}_{i=0}^{N-1}$,
 $\bu_k$ the predicted control sequence $\{u_{i|k}\}_{i=0}^{N-1}$,
${\bf c}_k$ the predicted control perturbations $\{c_{i|k}\}_{i=0}^{N-1}$,
$\bw_k$ the disturbance sequence $\{w_{i|k}\}_{i=0}^{N-1}$,
$\bv_k$ the sensor noise sequence $\{v_{i|k}\}_{i=0}^{N-1}$, and
$\bs{\zeta}_k$ the future innovation sequence $\{\gamma_{i|k} (y_{i|k}-C\xhat_{i|k}) \}_{i=0}^{N-1}$ at time instant $k$.
For a given sequence of matrices $\{\Psi_{i|k}\}_{i=0}^{N-1}$ and matrix $B$ let
\begin{alignat*}{2}
{\bf S}_{\Psi} &\!:=\!\! \begin{bmatrix} I \\ \Psi_{0|k} \\ \vdots \\ \prod\limits_{i=N-2}^0 \!\!\!\Psi_{i|k} \end{bmatrix}\!,
&~
{\bf T}_{(\Psi,B)} &\!:=\!\!
\begin{bmatrix}
0 & \cdots & & 0 \\
B & & &\\
\vdots & &\ddots & \\
\prod\limits_{i=N-2}^1 \!\!\!\Psi_{i|k} B \,\,~ & \cdots & B & 0
\end{bmatrix}\!,
\\
S^N_{\Psi} &:= \prod_{i=N-1}^0 \!\!\!\Psi_{i|k} \,,
&~
T^N_{(\Psi,B)} &:= \begin{bmatrix} \prod\limits_{i=N-1}^1 \!\!\!\Psi_{i|k} B \,~& \cdots & B \end{bmatrix},
\end{alignat*}
where $\prod\limits_{i=m}^n \Psi_{i|k} := \Psi_{m|k} \cdots \Psi_{n|k}$ for $m \geq n$,
and define
\begin{align*}
{\bf L}_k &:= \begin{bmatrix} L_{0,0|k} & & & \\ L_{1,0|k} & L_{1,1|k} & & \\ \vdots & \vdots & \ddots & \\ L_{N-1,0|k} & L_{N-1,1|k} & \cdots & L_{N-1,N-1|k} \end{bmatrix} ,\\
\bs{\Gamma}_k &:= \diag\{ \gamma_{0|k}, \ldots, \gamma_{N-1|k} \} \otimes I_{n_y \times n_y} ,
\end{align*}
${\bf K} := I_{N\times N}\otimes K$, ${\bf M} := I_{N\times N}\otimes M$ and ${\bf C} := I_{N\times N} \otimes C$.
Then from (\ref{eq:state_prediction}) we have
\begin{equation}\label{eq:pred_xdiff}
\bx_k \!- \bxhat_k \!= {\bf S}_{\Psi} (x_k - \xhat_k)
- {\bf T}_{(\Psi,A)} {\bf M}\bs{\Gamma}_k \bv_k
+ {\bf T}_{(\Psi,D)} \bw_k
\end{equation}
while (\ref{eq:nominal_state_estimate}) and (\ref{eq:predicted_control_law}b) give
\[
\bxhat_k = {\bf S}_{\Phi} \xhat_k + {\bf T}_{(\Phi,B)}({\bf c}_k + {\bf L}_k\bs{\zeta}_k) + {\bf T}_{(\Phi,A)} {\bf M} \bs{\zeta}_k .
\]
Here $\bs{\zeta}_k = \bs{\Gamma}_k {\bf C}(\bx_k - \bxhat_k) + \bs{\Gamma}_k\bv_k$ and ${\bf S}_{\Phi}$, ${\bf T}_{(\Phi,B)}$ are defined (analogously to ${\bf S}_{\Psi}$, ${\bf T}_{(\Psi,B)}$) in terms of $\Phi$ and $B$. Hence
\begin{subequations}\label{eq:pred_xhat_u}
\begin{align}
\hspace{-1mm}\bxhat_k &\!=\! {\bf S}_{\Phi}\xhat_k \!+\! {\bf T}_{(\Phi,B)}{\bf c}_k \!+\! \!({\bf T}_{(\Phi,B)}{\bf L}_k \!\!+\! {\bf T}_{(\Phi,A)}{\bf M}) \bs{\zeta}_k,
\\
\hspace{-1mm}\bu_k &\!=\! {\bf K} \bxhat_k \!+\! {\bf c}_k \!+\! {\bf L}_k \bs{\zeta}_k.
\end{align}
\end{subequations}
Clearly the predicted estimation error, state and control sequences in (\ref{eq:pred_xdiff}) and (\ref{eq:pred_xhat_u}a,b) depend linearly on the decision variables
$\theta_k \defeq ({\bf c}_k, {\bf L}_k)$.

\subsection{First and second moments of predicted sequences}

To express the cost and constraints of problem~(\ref{eq:opt}) in terms of the parameterisation introduced in Section~\ref{sec:control_parameterisation}, we derive in this section expressions for the means and variances of predicted state and control sequences.
%

%
From (\ref{eq:system}a) and \eqref{eq:cl loop priori estimate update}, it follows that 
\[
x_k - \xhat_k = \Psi_{k-1} (x_{k-1} - \xhat_{k-1}) - \gamma_{k-1} AM v_{k-1} + D w_{k-1}
\]
for all $k\geq 1$. Let $\Sigma_k$ denote the second moment of the state estimate error at time instant $k$:
\[
\Sigma_k \defeq \EE_k \bigl\{(x_k-\xhat_k)(x_k - \xhat_k)^\top \bigr\}.
\]
Then $\Sigma_k$ evolves according to
\begin{equation}\label{eq:cl_varx}
\Sigma_k \!=\! \Psi_{k\!-\!1} \Sigma_{k\!-\!1} \Psi_{k\!-\!1}^\top \!+\! \gamma_{k\!-\!1}AM\Sigma_v M^\top A^\top \!\!+\! D \Sigma_w D^\top
\end{equation}
for all $k\geq 1$, with initial condition $\Sigma_0$, and, by Assumption \ref{assumption:stabilising_gains}, $\EE \{\Sigma_k\}$ remains upper bounded $\forall k\geq 0$.

We can now derive the first and second moments of the predicted state sequence $\bx_k$ and control sequence $\bu_k$.

\begin{prop}\label{prop:moments_x_u}
Let $\bs{\pi}_k$, $\bs{\Pi}_k$, and $\bs{\Omega}_k$ be defined
\begin{gather*}
\bs{\pi}_k := {\bf S}_{\Phi} \xhat_k + {\bf T}_{(\Phi,B)}{\bf c}_k ,
~~
\bs{\Pi}_k := {\bf T}_{(\Phi,B)} {\bf L}_k + {\bf T}_{(\Phi,A)}{\bf M} ,
\\
\bs{\Omega}_k := \EE_k \biggl\{
\begin{bmatrix} \bx_k - \bxhat_k \\ \bs{\zeta}_k \end{bmatrix}
\begin{bmatrix} \bx_k - \bxhat_k \\ \bs{\zeta}_k \end{bmatrix}^\top
\biggr\}.
\end{gather*}
Then
\begin{subequations}\label{eq:1st_pred_x_u}
\begin{align}
&\EE_k \{\bx_k\}  = \EE_k\{\bxhat_k\} = \bs{\pi}_k,
\\
&\EE_k \{\bu_k\} = {\bf K} \EE_k\{\bxhat_k\} + {\bf c}_k
= {\bf K}\bs{\pi}_k + {\bf c}_k,
\end{align}
\end{subequations}
and
\begin{subequations}\label{eq:2nd_pred_x_u}
\begin{align}
&\!\!\EE_k\{\bx_k\bx_k^\top\} = \bs{\pi}_k\bs{\pi}_k^\top + \begin{bmatrix} I & \bs{\Pi}_k\end{bmatrix} \bs{\Omega}_k \begin{bmatrix} I \\ \bs{\Pi}_k^\top\end{bmatrix},
\\
&\!\!\EE_k\{\bu_k\bu_k^\top\} = ({\bf K}\bs{\pi}_k + {\bf c}_k)({\bf K}\bs{\pi}_k + {\bf c}_k)^\top
\nonumber\\
&\qquad\quad + \begin{bmatrix} 0 & {\bf L}_k+{\bf K}\bs{\Pi}_k\end{bmatrix} \bs{\Omega}_k \begin{bmatrix} 0 \\ ({\bf L}_k+{\bf K}\bs{\Pi}_k)^\top\end{bmatrix} .
\end{align}
\end{subequations}
\end{prop}
\vspace{-\baselineskip}
\bpf
From~(\ref{eq:pred_xdiff}), we have $\EE_k \{\bx_k - \bxhat_k \} = 0$. Therefore $\bs{\zeta}_k = \bs{\Gamma}_k{\bf C}(\bx_k - \bxhat_k) + \bs{\Gamma}_k{\bf v}_k$ implies $\EE_k\{\bs{\zeta}_k \} = 0$
and (\ref{eq:1st_pred_x_u}a,b) follow from the expectations of (\ref{eq:pred_xhat_u}a,b).
To determine the second moments of $\bx_k$ and $\bu_k$, let
\[
{\bf X}_k \defeq \EE_k\biggl\{
\begin{bmatrix} \bx_k - \bxhat_k \\ \bxhat_k \end{bmatrix}
\begin{bmatrix} \bx_k - \bxhat_k \\ \bxhat_k \end{bmatrix}^\top
\biggr\}\ .
\]
Then from (\ref{eq:pred_xdiff}) and (\ref{eq:pred_xhat_u}a) we have
\begin{equation}\label{eq:x_covar}
{\bf X}_k =
\begin{bmatrix} 0 & 0 \\ 0 & \bs{\pi}_k \bs{\pi}_k^\top \end{bmatrix}
+
\begin{bmatrix} I & 0 \\ 0 & \bs{\Pi}_k\end{bmatrix} \bs{\Omega}_k \begin{bmatrix} I & 0 \\ 0 & \bs{\Pi}_k\end{bmatrix}^\top ,
\end{equation}
and (\ref{eq:2nd_pred_x_u}a,b) follow from
$\EE_k\{\bx_k \bx_k^\top \} = \bigl[ I \ I \bigr] {\bf X}_k \bigl[ I \ I \bigr]^\top$
and (\ref{eq:pred_xhat_u}a,b), respectively.
\epf

Since $\bs{\pi}_k$, $\bs{\Pi}_k$ are linear in $\theta_k \!=\! ({\bf c}_k,{\bf L}_k)$ and $\bs{\Omega}_k$ is independent of $\theta_k$ (as will be shown by \eqref{eq:cov_mat}),
it is clear from (\ref{eq:1st_pred_x_u}a,b) and (\ref{eq:2nd_pred_x_u}a,b)
that the first moments of the predicted state and input sequences are linear in $\theta_k$ while their second moments are quadratic functions of $\theta_k$.

To determine $\bs{\Omega}_k$, note that $\bx_k - \bxhat_k$ and $\bs{\zeta}_k$ can be written
\[
\bx_k - \bxhat_k = F(\bs{\Gamma}_k)q_k ,
~~
\bs{\zeta}_k = G(\bs{\Gamma}_k)q_k  ,
~~
q_k := \begin{bmatrix} x_k - \hat{x}_k \\ {\bf v}_k \\ {\bf w}_k \end{bmatrix},
\]
with
$F(\bs{\Gamma}_k) := \begin{bmatrix} {\bf S}_{\Psi} & {-{\bf T}_{(\Psi,A)}}{\bf M} \bs{\Gamma}_k & {\bf T}_{(\Psi,D)} \end{bmatrix}$ and
$G(\bs{\Gamma}_k) := \bs{\Gamma}_k{\bf C}F(\bs{\Gamma}_k) + \begin{bmatrix} 0 & \bs{\Gamma}_k & 0 \end{bmatrix}$. Hence, by the law of total expectation,
\begin{equation}\label{eq:cov_mat}
\bs{\Omega}_k \!=\!\!
\sum_j \!\begin{bmatrix} F(\bs{\Gamma}^{(j)}) \\ G(\bs{\Gamma}^{(j)}) \end{bmatrix} \!\EE_k\{ q_k q_k^\top \}\! \begin{bmatrix} F(\bs{\Gamma}^{(j)}) \\ G(\bs{\Gamma}^{(j)}) \end{bmatrix}^\top \!\!\PP \{ \bs{\Gamma}_k \!=\! \bs{\Gamma}^{(j)} \} ,
\end{equation}
where $\EE_k\{ q_k q_k^\top \}$ is the block-diagonal matrix:
\begin{gather*}
\EE_k\{ q_k q_k^\top \} = \diag\{ \Sigma_k, \bar{\Sigma}_v, \bar{\Sigma}_w\},
\\
\bar{\Sigma}_v :=I_{N \times N} \otimes \Sigma_v,
\quad
\bar{\Sigma}_w :=I_{N \times N} \otimes \Sigma_w,
\end{gather*}
and where $\bs{\Gamma}^{(j)}$ for $j \!=\!1,\ldots,2^N$ enumerates the $2^N$ matrices with binary-valued diagonal elements defined by
\begin{align*}
&\bs{\Gamma}^{(1)} = 0, ~~
\bs{\Gamma}^{(2)} = \diag\{0,\ldots,0,1\}\otimes I_{n_y\times n_y},  ~~ \ldots
\\
&
\ldots, ~ \bs{\Gamma}^{(2^N-1)} = \diag\{1,\ldots,1 ,0\}\otimes I_{n_y\times n_y}, ~~
\bs{\Gamma}^{(2^N)} = I .
\end{align*}

\begin{rem}
The matrix \,$\bs{\Omega}_k $ in (\ref{eq:cov_mat}) can be computed conveniently via
\begin{align*}
&\vvec(\bs{\Omega}_k) =
\\
&\Bigl( \sum_j\!\!
\begin{bmatrix} F(\bs{\Gamma}^{(j)}) \\ G(\bs{\Gamma}^{(j)}) \end{bmatrix} \!\!\otimes\!\!
\begin{bmatrix} F(\bs{\Gamma}^{(j)}) \\ G(\bs{\Gamma}^{(j)}) \end{bmatrix}
\!\!\PP \{ \bs{\Gamma}_k \!=\! \bs{\Gamma}^{(j)} \} \!\Bigr)
\!\vvec\Bigl(\Bigl[\begin{smallmatrix} \!\Sigma_k\! & & \\ & \!\bar{\Sigma}_v\! & \\ & & \!\bar{\Sigma}_w\! \end{smallmatrix} \Bigr]\Bigr)
\end{align*}
where the first term on the RHS can be determined offline given the probability distribution of $\gamma_k$. This allows $\bs{\Omega}_k$ to be computed online using the current value of $\Sigma_k$ with a single matrix-vector multiplication.
\end{rem}


Using the same arguments as the proof of Proposition~\ref{prop:moments_x_u},
it can be verified that
\begin{align}
\hspace{-2.8mm}X_{N|k} &\!:=\!
\EE_k \biggl\{\begin{bmatrix} x_{N|k} \!-\! \xhat_{N|k} \\ \xhat_{N|k} \end{bmatrix}
\begin{bmatrix} x_{N|k} \!-\! \xhat_{N|k} \\ \xhat_{N|k} \end{bmatrix}^\top
\biggr\} \nonumber \\
&\!=\!
\begin{bmatrix} 0 & 0 \\ 0 & \pi_{N|k} \pi_{N|k}^\top \end{bmatrix}
\!+\!
\begin{bmatrix} I & 0 \\ 0 & \Pi_{N|k}\end{bmatrix} \Omega_{N|k} \begin{bmatrix} I & 0 \\ 0 & \Pi_{N|k}\end{bmatrix}^\top
\nonumber
\end{align}
where
\begin{align}
\Pi_{N|k} \!&:=\! T^N_{(\Phi,B)} {\bf L}_k + T^N_{(\Phi,A)} {\bf M},\,
\pi_{N|k} \!:=\! S^N_{\Phi} \xhat_k + T^N_{(\Phi,B)} {\bf c}_k, \nonumber
\\
\Omega_{N|k} \!&=\!\! \sum_j\! \begin{bmatrix} F_N(\bs{\Gamma}^{(j)}) \\ G (\bs{\Gamma}^{(j)}) \end{bmatrix}
\!\!\Bigl[ \begin{smallmatrix}
\!\Sigma_k \! & & \\ & \!\bar{\Sigma}_v\!\! & \\  & & \bar{\Sigma}_w\!
\end{smallmatrix} \Bigr]\!\!
\begin{bmatrix} F_N (\bs{\Gamma}^{(j)}) \\ G (\bs{\Gamma}^{(j)}) \end{bmatrix}^{\!\top}\!\! \!\PP \{ \bs{\Gamma}_k \!\!=\!\! \bs{\Gamma}^{(j)} \!\} \nonumber 
\end{align}
with
$F_N(\bs{\Gamma}_k) := \begin{bmatrix} S^N_{\Psi} & -T^N_{(\Psi,A)}{\bf M} \bs{\Gamma}_k & T^N_{(\Psi,D)} \end{bmatrix}$
and $S_{\Phi}^N$, $T_{(\Phi,A)}^N$, $T_{(\Phi,B)}^N$  being defined (analogously to $S_{\Psi}^N$ and $T^N_{(\Psi,B)}$) in terms of $\Phi$, $A$ and $B$.

\section{MPC optimisation} \label{sec:cost and constraints}
In this section, we formulate the MPC optimisation problem to be repeatedly solved online, using the expressions of the first and second moments of the predicted state and control sequences derived in Section \ref{sec:MPC_predictions}.

First note that the objective in (\ref{eq:opt}) can be written
\begin{multline}
\hspace{-2pt}\sum_{i=0}^{\infty} \beta^i \EE_k \bigl\{ \| x_{i|k} \|_Q^2 + \| u_{i|k}\|_R^2 \bigr\}
=
\trloc (\Q_\beta{\bf X}_k) + \trloc(\R_\beta{\bf U}_k)
\\
+ f_N(\theta_k,\xhat_k,\Sigma_k)
\label{eq:cost_bound}
\end{multline}
where ${\bf X}_k$ is given by (\ref{eq:x_covar}), and
\begin{align*}
&\Q_\beta  \defeq
\ones_{2\times 2} \!\otimes\!\diag\{ Q , \beta Q, \ldots, \beta^{N-1}Q \} ,
\\
&\R_\beta  \defeq \diag\{ R , \beta R, \ldots, \beta^{N-1}R \} ,
{\bf U}_k \!\defeq\! \EE_k \{\bu_k\bu_k^\top \},
\\
&f_N(\theta_k, \xhat_k,\Sigma_k) \defeq \sum_{i=N}^{\infty} \beta^i \EE_k \bigl\{ \| x_{i|k} \|_Q^2 + \| u_{i|k}\|_R^2 \bigr\} .
\end{align*}
Since $\Q_\beta\succeq 0$ and $\R_\beta \succ 0$, the term $\trloc (\Q_\beta{\bf X}_k) + \trloc(\R_\beta{\bf U}_k)$ in (\ref{eq:cost_bound}) can be expressed as a convex quadratic function of $\theta_k = ({\bf c}_k,{\bf L}_k)$ using (\ref{eq:2nd_pred_x_u}b) and (\ref{eq:x_covar}).
%
%
To determine the terminal term, $f_N(\theta,\xhat_k,\Sigma_k)$, let
\begin{equation}
P_k \defeq \sum_{i=N}^\infty \beta^i X_{i|k}, \label{eq:def_Pk}
\end{equation}
where
\[
X_{i|k} := \EE_k \biggl\{\begin{bmatrix} x_{i|k} - \xhat_{i|k} \\ \xhat_{i|k} \end{bmatrix}
\begin{bmatrix} x_{i|k} - \xhat_{i|k} \\ \xhat_{i|k} \end{bmatrix}^\top\biggr\} .
\]
Then for $i\geq N$ we have
\begin{equation} \label{eq:recursion X_i}
X_{i+1|k} \!=\! \EE \bigl\{ \tilde{\Psi}(\gamma) X_{i|k} \tilde{\Psi}^\top(\gamma) \bigr\}
+ \EE \bigl\{ \tilde{D}(\gamma) \bigl[\begin{smallmatrix}\Sigma_v & \\ & \Sigma_w\end{smallmatrix}\bigr]
\tilde{D}^\top (\gamma) \bigr\}
\end{equation}
where
\[
\tilde{\Psi}(\gamma) := \begin{bmatrix} A(I - \gamma MC) & 0 \\ \gamma A M C & \Phi \end{bmatrix} ,
\quad
\tilde{D}(\gamma) := \begin{bmatrix} -\gamma A M & D \\ \gamma A M & 0\end{bmatrix} ,
\]
and $\gamma$ is essentially $\gamma_{i|k}$ with subscripts being omitted for simplicity.
Hence
\begin{align*}
& \EE\bigl\{ \tilde{\Psi}(\gamma) P_k \tilde{\Psi}^\top(\gamma) \bigr\} \\
&=
\sum_{i=N}^\infty \beta^i \bigl( X_{i+1|k} - \EE\{\tilde{D}(\gamma) \bigl[\begin{smallmatrix}\Sigma_v & \\ & \Sigma_w\end{smallmatrix}\bigr]
\tilde{D}^\top (\gamma) \}\bigr)
\\
&= \beta^{-1} ( P_k \!-\! \beta^N X_{N|k}) -\!\frac{\beta^N}{1-\beta} \EE \{\tilde{D}(\gamma)
\bigl[\begin{smallmatrix}\Sigma_v & \\ & \Sigma_w\end{smallmatrix}\bigr]
\tilde{D}^\top (\gamma) \} ,
\end{align*}
and the terminal term $f_N(\theta_k,\xhat_k,\Sigma_k)$ in (\ref{eq:cost_bound}) is equal to $\trloc( Z_1 P_k )$
where $Z_1 \defeq \Bigl[\begin{smallmatrix}  Q & ~Q \\ Q & ~Q + K^\top\! RK \end{smallmatrix}\Bigr] $ and
$P_k$ is the solution to the stochastic Lyapunov equation
\begin{align}
P_k &=\beta \EE\bigl\{ \tilde{\Psi}(\gamma) P_k \tilde{\Psi}^\top(\gamma) \bigr\}
\nonumber \\
&\quad + \beta^N X_{N|k} + \frac{\beta^{N+1}}{1-\beta} \EE\bigl\{ \tilde{D}(\gamma)
\Bigl[\begin{smallmatrix}\!\Sigma_v\! & \\ & \!\Sigma_w\! \end{smallmatrix}\Bigr]
\tilde{D}^\top(\gamma) \bigr\} .
\label{eq:term_cost_sdp}
\end{align}
\begin{lem} \label{lemma:MSS_tPsi}
By Assumption \ref{assumption:stabilising_gains}, the linear system $\xi_{i+1}=\tilde{\Psi}(\gamma)\xi_i$ is mean-square stable.
\end{lem}
\vspace{-\baselineskip}
\bpf
In Assumption \ref{assumption:stabilising_gains}, it is assumed that
$\Psi(\gamma):=A(I-\gamma MC)$ is MSS and $\Phi$ is asymptotically stable, which are equivalent to
\begin{align*}
\exists~ \Xi_1=\Xi_1^\top \succ 0, \Xi_1 -\EE\{ \Psi(\gamma)\Xi_1 \Psi^\top(\gamma)\} \succ 0 ,\\
\exists~ \Xi_2=\Xi_2^\top \succ 0, \Xi_2 - \Phi \Xi_2 \Phi^\top \succ 0,
\end{align*}
respectively. It can be shown that $\Bigl[\begin{smallmatrix} \Xi_1 & \\ &\Xi_2 \end{smallmatrix}\Bigr]$ satisfies
\[
\Bigl[\begin{smallmatrix} \Xi_1 & \\ &\Xi_2 \end{smallmatrix}\Bigr] -\EE\{ \tilde{\Psi}(\gamma) \Bigl[\begin{smallmatrix} \Xi_1 & \\ &\Xi_2 \end{smallmatrix}\Bigr] \tilde{\Psi}^\top(\gamma) \}\succ 0,
\]
and this implies Lemma \ref{lemma:MSS_tPsi}.
\epf

Let $\bar{X}\succeq 0$ denote the steady state solution to \eqref{eq:recursion X_i}. Then,
by Lemma \ref{lemma:MSS_tPsi}, it is ensured that $X_{i|k}$ converges to $\bar{X}$ as $i\to \infty$ for any $X_{N|k}\succeq 0$ and $k \geq 0$. Also, Lemma \ref{lemma:MSS_tPsi} implies $\beta^{\frac{1}{2}} \tilde{\Psi}(\gamma)$ is MSS since $\beta\in (0,1)$ and thus $P_k$ in \eqref{eq:def_Pk} is well defined and finite.

Re-writing the constraints of problem (\ref{eq:opt}) using the matrix, $X_{i|k}$, of second moments yields the condition
\[
\sum_{i=0}^\infty
\beta^i
\trloc\bigl[ ( \ones_{2\times 2}
\otimes H^\top H)  X_{i|k}\bigr]
\leq \mu_{k},
\]
which is equivalent to the constraint
\begin{equation}\label{eq:discounted_constraint}
\trloc({\bf H}_\beta {\bf X}_k)
+
\trloc( Z_2  P_k)
\leq \mu_{k},
\end{equation}
where ${\bf H}_\beta \!:=\! \ones_{2\times 2} \otimes \diag \{ H^\top \!H, \beta H^\top \!H, \ldots, \beta^{N-1} H^\top \!H \}$
and $Z_2:= \ones_{2\times 2} \otimes H^\top H$.

The expressions for the cost and constraints in (\ref{eq:cost_bound}), \eqref{eq:term_cost_sdp}, (\ref{eq:discounted_constraint}) allow the optimisation~(\ref{eq:opt}) defining $\theta_k^\ast$ to be formulated as
\begin{equation} \label{eq:opt2}
\begin{aligned}
(\theta^\ast_k,P_k^\ast) := \arg\min_{\theta_k,P_k}
& \trloc (\Q_\beta{\bf X}_k) + \trloc(\R_\beta{\bf U}_k) + \trloc(Z_1 P_k)
\\
\text{s.t.} \
& \begin{aligned}[t]
&
\eqref{eq:discounted_constraint},
\\
& \!P_k \succeq \beta \EE\bigl\{ \tilde{\Psi}(\gamma) P_k \tilde{\Psi}^\top(\gamma) \bigr\} \!+\! \beta^N X_{N|k} \\
& ~\quad+ \frac{\beta^{N+1}}{1-\beta} \EE\bigl\{ \tilde{D}(\gamma) \Bigl[\begin{smallmatrix} \!\Sigma_v\! & \\ & \!\Sigma_w\!\end{smallmatrix} \Bigr] \tilde{D}^{\top}\!(\gamma) \bigr\} .
\end{aligned}
\end{aligned}
\end{equation}
\begin{rem} \label{remark:relaxation to lmi nonconservative}
The relaxation of
(\ref{eq:term_cost_sdp}) as a linear matrix inequality (LMI) in problem \eqref{eq:opt2} does not introduce any conservativeness as it can be shown that there always exists a solution that satisfies this LMI with equality. Suppose there does not exist such a solution to problem \eqref{eq:opt2}. Let
$(\theta^\ast_k,P_k^{(0)})$ be a minimiser and
\begin{multline*}
P_k^{(1)} := \beta \EE\bigl\{ \tilde{\Psi}(\gamma) P_k^{(0)} \tilde{\Psi}^\top(\gamma) \bigr\}+ \beta^N   X_{N|k} \\
+\frac{\beta^{N+1}}{1-\beta} \EE\bigl\{ \tilde{D}(\gamma) \Bigl[\begin{smallmatrix} \!\Sigma_v\! & \\ & \!\Sigma_w\!\end{smallmatrix} \Bigr] \tilde{D}^{\top}\!(\gamma) \bigr\}.
\end{multline*}
Then $P_k^{(0)} \succ P_k^{(1)}$ and $(\theta^\ast_k,P_k^{(1)})$ also satisfies the LMI with strict inequality because of the supposition. Continuing this procedure, we will have that $\,\forall i \geq 1 $ $P_k^{(i)} \succ P_k^{(i+1)}$ and that $(\theta^\ast_k,P_k^{(i+1)})$ is a minimiser and is at least as good as $(\theta^\ast_k,P_k^{(i)})$ in terms of values of the cost function. From the mean-square stability of $\beta^{\frac{1}{2}}\tilde{\Psi}(\gamma)$, it follows that $P_k^\ast = \lim_{i \to \infty} P_k^{(i)}$ exists and satisfies this LMI with equality. This contradicts the supposition and proves our argument. Therefore, we let $(\theta^\ast_k,P_k^\ast)$ be the minimiser that satisfies this LMI constraint with equality.
Note that if
$\Bigr[\begin{smallmatrix} Q^{\frac{1}{2}} & 0 \\ Q^{\frac{1}{2}} & K^\top R^{\frac{1}{2}} \end{smallmatrix}\Bigl]$ has full column rank, $(\theta_k^\ast,P_k^\ast)$ is unique.
\end{rem}

\section{Closed loop properties} \label{sec:properties}

This section considers the performance of the closed loop system~(\ref{eq:system}) with the control law of Section~\ref{sec:control_implementation}.
We use the solution $\theta^\ast_k=\{{\bf c}^\ast_k,{\bf L}^\ast_k\}$ of (\ref{eq:opt}) at time instant $k$ to construct a feasible, but possibly suboptimal, solution for (\ref{eq:opt}) at time instant \mbox{$k+1$} (i.e.~given $\U_k$, $\I_k$), which we denote  $\theta^\tail_{k+1}:= \{{\bf c}^\tail_{k+1},{\bf L}^\tail_{k+1}\}$, where
\begin{subequations}\label{eq:tail}
\begin{align}
{\bf c}^\tail_{k+1} &\defeq
\mathcal{T} {\bf c}_k^\ast
+
\widetilde{\bf L}^\ast_{0|k} \gamma_k (y_k - C\xhat_k )  ,
\label{eq:c_tail}
\\
{\bf L}^\tail_{k+1} &\defeq
\begin{bmatrix}
L^\ast_{1,1|k} & & & \\
\vdots & \ddots & & \\
L^\ast_{N-1,1|k} & \cdots & L^\ast_{N-1,N-1|k} & \\
0 & \cdots & 0 & 0
\end{bmatrix} .
\end{align}
\end{subequations}
In \eqref{eq:c_tail}, $\mathcal{T}$ is a matrix such that
\[
\mathcal{T}
\begin{bmatrix}
c_0 \\ c_1 \\\vdots \\  c_{N-1} 
\end{bmatrix}
=
\begin{bmatrix}
c_1 \\ \vdots \\  c_{N-1} \\ 0
\end{bmatrix},~\text{and}~
\widetilde{\bf L}^\ast_{0|k}:=
\begin{bmatrix}
L^\ast_{1,0|k} \\
 \vdots \\
L^\ast_{N-1,0|k} \\
0
\end{bmatrix}.
\]
Following \cite{yan:18}, we define the constraint threshold $\mu_{k}$ in~(\ref{eq:opt}) $\forall k>0$ in terms of $\theta^\tail_{k}$ 
as
\begin{equation}\label{eq:mu_def}
\mu_k \defeq \begin{cases} \epsilon , & k = 0
\\
\trloc ( {\bf H}_\beta {\bf X}^\tail_{k}) +
\trloc ( Z_2  P^\tail_k ) ,  & k > 0
\end{cases}
\end{equation}
where
\begin{subequations} \label{eq:feasible Pk and X_k}
\begin{align}
&P^\tail_k := \beta \EE\bigl\{ \tilde{\Psi}(\gamma) P^\tail_k \tilde{\Psi}^\top(\gamma) \bigr\} \nonumber
\\
& \quad + \beta^N X^\tail_{N|k} + \frac{\beta^{N+1}}{1-\beta} \EE\bigl\{ \tilde{D}(\gamma) \Bigl[\begin{smallmatrix} \Sigma_v & \\ & \Sigma_w\end{smallmatrix} \Bigr] \tilde{D}^{\top}(\gamma) \bigr\},
\\
&{\bf X}^\tail_k :=
\begin{bmatrix} 0 & 0 \\ 0 & \bs{\pi}^\tail_k \bs{\pi}_k^{\tail\,\top} \end{bmatrix}
+
\begin{bmatrix} I & 0 \\ 0 & \bs{\Pi}_k^\tail\end{bmatrix} \bs{\Omega}_k \begin{bmatrix} I & 0 \\ 0 & \bs{\Pi}_k^\tail\end{bmatrix}^\top,
\\
&X_{N|k}^\tail :=
\begin{bmatrix} 0 & 0 \\ 0 & \pi_{N|k}^\tail \pi_{N|k}^{\tail\,\top} \end{bmatrix}
+
\begin{bmatrix} I & 0 \\ 0 & \Pi_{N|k}^\tail\end{bmatrix} \Omega_{N|k} \begin{bmatrix} I & 0 \\ 0 & \Pi_{N|k}^\tail\end{bmatrix}^\top
\end{align}
\end{subequations}
with
\begin{align*}
&\bs{\pi}_k^\tail := {\bf S}_{\Phi} \xhat_k \!+\! {\bf T}_{(\Phi,B)}{\bf c}_k^\tail, ~
\bs{\Pi}_k^\tail := {\bf T}_{(\Phi,B)} {\bf L}_k^\tail \!+\! {\bf T}_{(\Phi,A)}{\bf M},
\\
&\pi_{N|k}^\tail := {S}^N_{\Phi} \xhat_k \!+\! {T}^N_{(\Phi,B)}{\bf c}_k^\tail,\,
{\Pi}_{N|k}^\tail := {T}^N_{(\Phi,B)} {\bf L}_k^\tail \!+\! {T}^N_{(\Phi,A)}{\bf M}.
\end{align*}
Combining \eqref{eq:discounted_constraint} and \eqref{eq:mu_def}, we can see that the design of $\mu_k$ enforces feasibility of $\theta^\tail_{k}$ $\forall k>0$ and therefore 
ensures recursive feasibility of the MPC optimisation without requiring bounds on the noise $v_k$ and disturbance $w_k$.

\begin{thm}\label{thm:cl_constraint}
If problem~\eqref{eq:opt2} is feasible at $k=0$, then \eqref{eq:opt2} remains feasible for all $k>0$ and the state of (\ref{eq:system}) under the control law of Section~\ref{sec:control_implementation} satisfies
\begin{equation}\label{eq:cl_constraint}
\sum_{k=0}^\infty \beta^k \EE \{ \| H x_{k}\|^2\} \leq \epsilon  .
\end{equation}
\end{thm}
\vspace{-\baselineskip}
\bpf
The definition (\ref{eq:mu_def}) of $\mu_k$ trivially ensures feasibility for all $k>0$.
The definitions (\ref{eq:tail}a,b) ensure that,
at time instant $k$ (given $\U_{k-1}$, $\I_{k-1}$),
the distributions of the state and control sequences $\{x_{i|k+1}\}_{i=0}^\infty$ and $\{u_{i|k+1}\}_{i=0}^\infty$ are identical to the distributions of $\{x_{i+1|k}\}_{i=0}^\infty$ and $\{u_{i+1|k}\}_{i=0}^\infty$.
%
%
Therefore
\[
\sum_{i=0}^\infty \beta^i \EE_k \{ \|H x_{i|k}\|^2 \}
=
\trloc({\bf H}_\beta {\bf X}_k)
+
\trloc ( Z_2  P_k )
\]
implies
\begin{equation} \nonumber 
\beta \EE_k\{ \mu_{k+1} \}
\leq \mu_k - \EE_k\{ \|Hx_{0|k}\|^2 \}
= \mu_k - \EE_k\{ \|Hx_{k}\|^2 \} .
\end{equation}
Hence the trajectories of the closed loop system satisfy
\[
\sum_{i=0}^\infty \beta^{i} \EE_k \{ \|H x_{k+i}\|^2 \}
\leq
\mu_k - \lim_{i\to\infty}\beta^i \EE_{k} \{ \mu_{k+i} \}
\leq
\mu_k
\]
for all $k \geq 0$.
\epf

\begin{cor}\label{cor:cl_cost}
Let $J_k:=J(\theta_k^\ast,\xhat_k,\Sigma_k)$ denote the optimal cost of problem \eqref{eq:opt2}. Then, under the control law of Section~\ref{sec:control_implementation}, the trajectories of (\ref{eq:system})  satisfy
\begin{equation}\label{eq:cl_cost}
\sum_{k=0}^{\infty} \beta^k \EE \bigl\{ \| x_{k} \|_Q^2  + \| u_{k}\|_R^2 \bigr\} \leq J_0 .
\end{equation}
\end{cor}
\vspace{-\baselineskip}
\bpf
Using the same argument in the proof of Theorem~\ref{thm:cl_constraint}
on the definition of the objective in \eqref{eq:opt2} yields
\begin{equation}
\beta \EE_k \!\{\! J(\theta^\tail_{k+1},\xhat_{k+1},\Sigma_{k+1})\!\}
\!\!=\!\! J_k \!-\! \EE_k\!\{\|x_{k}\|_Q^2 \!+\! \|u_{k}\|_R^2\} , \label{eq:equality_cl_cost}
\end{equation}
and since $ J_k \leq J(\theta^\tail_{k},\xhat_{k},\Sigma_{k}) ~\forall k$ by optimality, the bound in (\ref{eq:cl_cost}) follows.
\epf

\section{Strengthened stability conditions} \label{sec:strengthened stability}
In Corollary \ref{cor:cl_cost}, we provide an upper bound on the discounted closed loop cost. However, this may not prevent closed loop states from gradually diverging
if the closed loop cost is heavily discounted. Therefore, in the following,
we provide stability conditions in terms of bounds on the  undiscounted closed loop cost.

\subsection{Closed loop cost bound via a numerical check}\label{subsec:upper bounds on cost}
In this section, we provide a bound on the averaged undiscounted closed loop cost accumulated over an infinite horizon
by running a numerical check on $\beta$. Specifically, if some condition holds for values of $\beta$ in $(0,1)$, we can provide a bound on this closed loop cost that is parametric in $\beta$. To this end, we first establish an upper bound on $J_k$.

An upper bound on $J_k$ can be provided by finding a feasible solution to problem \eqref{eq:opt2}, and such a solution can be easily obtained if we eliminate variable $P_k$ and rewrite problem \eqref{eq:opt2} as follows. Using standard matrix vectorisation identities, \eqref{eq:term_cost_sdp} can be rewritten as
\begin{equation}
\vvec(P_k) = W_1 \vvec(\beta^N X_{N|k} + W_2), \label{eq:vec_Pk}
\end{equation}
where
\begin{align*}
W_1 \defeq & \bigl[ I - \beta (1-\lambda) \tilde{\Psi}(0)\otimes\tilde{\Psi}(0) - \beta\lambda\tilde{\Psi}(1)\otimes\tilde{\Psi}(1)\bigr]^{-1}, \\
W_2 \defeq & \frac{\beta^{N+1}}{1-\beta}
\EE\bigl\{ \tilde{D}(\gamma)
\Bigl[\begin{smallmatrix}\Sigma_v & \\ & \Sigma_w \end{smallmatrix}\Bigr]
\tilde{D}^\top(\gamma) \bigr\}.
\end{align*}
In the definition of $W_1$, the matrix inverse is guaranteed to exist since $\beta^{\frac{1}{2}} \tilde{\Psi}(\gamma)$ is MSS.
From \eqref{eq:vec_Pk} we have
\begin{multline}
\trloc (Z_1 P_k)=\vvec(Z_1^\top)^\top \vvec(P_k)\\
\begin{aligned}
&=\vvec(Z_1)^\top W_1 \vvec(\beta^N X_{N|k} + W_2) \\
&=\trloc (\beta^N \tilde{Z}_1  X_{N|k}) + \trloc (\tilde{Z}_1 W_2),
\end{aligned}  \nonumber
\end{multline}
where $\tilde{Z}_1^\top \defeq \reshape\bigl( (\vvec(Z_1)^\top W_1)^\top, [2n_x,2n_x] \bigr)$.
Similarly, 
$
\trloc( Z_2  P_k)
\!=\!\trloc (\beta^N \tilde{Z}_2 X_{N|k}) + \trloc (\tilde{Z}_2 W_2),
$
where $\tilde{Z}_2^\top \defeq \reshape \bigl( (\vvec(Z_2)^\top W_1)^\top, [2n_x,2n_x] \bigr)$.

Therefore, an equivalent form of problem \eqref{eq:opt2} is given by
\begin{gather}
\theta^\ast_k = \arg\min_{\theta_k}
\trloc (\Q_\beta{\bf X}_k) \!+\! \trloc (\R_\beta{\bf U}_k) \!+\!
\beta^N \trloc ( \tilde{Z}_1  X_{N|k}) \nonumber \\
\hspace{6cm}+ \trloc (\tilde{Z}_1 W_2) \nonumber
\\
\!\!\text{s.t.} \  \trloc ({\bf H}_\beta {\bf X}_k)
\!+\!
\beta^N  \trloc (\tilde{Z}_2 X_{N|k}) \!+\! \trloc (\tilde{Z}_2 W_2)
\leq \mu_{k}, \label{eq:opt3}
\end{gather}
which has a single convex quadratic constraint.

Minimising the constraint function in \eqref{eq:opt3} over $\theta_k=\{{\bf c}_k, {\bf L}_k\}$ necessarily yields a feasible, but possibly suboptimal, solution $\theta^f_k:=\{{\bf c}^f_k, {\bf L}^f_k\}$. This solution has an explicit form, in which ${\bf c}_k^f$ is given by ${\bf c}_k^f:=K_c \xhat_k$ with
\begin{multline}
\hspace{-2mm}K_c \defeq -\bigl[ {\bf T}^\top_{(\Phi,B)} H_b {\bf T}_{(\Phi,B)}+  \beta^N(T^N_{(\Phi,B)})^\top \tilde{Z}_{2,22}^\top  T^N_{(\Phi,B)} \bigr]^\dagger \\ \cdot \bigl[ {\bf T}^\top_{(\Phi,B)} H_b {\bf S}_{\Phi} + \frac{\beta^N}{2}(T^N_{(\Phi,B)})^\top (\tilde{Z}_{2,22}^\top+\tilde{Z}_{2,22}) S^N_{\Phi}\bigr], \nonumber
\end{multline}
$H_b \defeq \diag \{ H^\top H, \beta H^\top H, \ldots, \beta^{N-1} H^\top H \}$.
The matrix
$\tilde{Z}_{2,22}$ is the bottom-right block of $\tilde{Z}_2$ partitioned according to
$\tilde{Z}_2=\Bigl[\begin{smallmatrix} \tilde{Z}_{2,11} & \tilde{Z}_{2,12} \\
\tilde{Z}_{2,21}& \tilde{Z}_{2,22} \end{smallmatrix}\Bigr]$.
The expression for ${\bf L}_k^f$ is omitted here for simplicity but we note that it is a function of $\Sigma_k$.
Therefore, given $\theta_k^f=\{{\bf c}_k^f,{\bf L}_k^f\}$ we can express the cost function as
\begin{equation}
J(\theta_k^f,\xhat_k,\!\Sigma_k)\!=\!\xhat^\top_k P_{\xhat} \xhat_k + \trloc(Q\Sigma_k)+ s(\beta,\!\Sigma_k) + \eta. \label{eq:J_of_theta_k^f}
\end{equation}
In \eqref{eq:J_of_theta_k^f}, $s(\beta,\Sigma_k)$ is a nonnegative scalar function of $\beta$ and $\Sigma_k$, consisting of some terms in $J(\theta_k^f,\xhat_k,\Sigma_k)$ that depend on $\Sigma_k$ and taking finite values when $\Sigma_k$ is finite, and $\eta$ is a constant depending on $\beta$, $\Sigma_v$, $\Sigma_w$ and other system parameters. The matrix $P_{\xhat}$ is given by
\begin{multline*}
P_{\xhat} \defeq ({\bf S}_{\Phi}+{\bf T}_{(\Phi,B)}K_c)^\top Q_b ({\bf S}_{\Phi}+{\bf T}_{(\Phi,B)}K_c) \\
+ ({\bf K}({\bf S}_{\Phi}\!+\!{\bf T}_{(\Phi\!,B)}K_c)\!+\!K_c)^\top \R_\beta ({\bf K}({\bf S}_{\Phi}\!+\!{\bf T}_{(\Phi\!,B)}K_c)\!+\!K_c) \\
\hspace{1pt}+ \frac{\beta^N}{2}(S_\Phi^N+T_{(\Phi,B)}^NK_c)^\top (\tilde{Z}_{1,22}^\top +\tilde{Z}_{1,22} ) (S_\Phi^N+T_{(\Phi,B)}^NK_c)
\end{multline*}
where $Q_b \defeq \diag\{ Q , \beta Q, \ldots, \beta^{N-1}Q \} $ and $\tilde{Z}_{1,22}$ is the bottom-right block of $\tilde{Z}_1=\Bigl[\begin{smallmatrix}\tilde{Z}_{1,11}&\tilde{Z}_{1,12} \\\tilde{Z}_{1,21}&\tilde{Z}_{1,22} \end{smallmatrix}\Bigr]$.

Moreover, if $\sigma$ is any scalar such that
\begin{equation}
P_{\xhat} \preceq \sigma Q, \label{eq:condition on sigma}
\end{equation}
then we have
\begin{align}
J_k &\leq J(\theta_k^f,\xhat_k,\Sigma_k) \nonumber \\
&\leq \sigma \trloc( Q \xhat_k \xhat^\top_k ) + \trloc(Q\Sigma_k) + s(\beta,\Sigma_k)+ \eta \nonumber
\\
&\leq \sigma \EE_k \{ \norm{x_k}^2_Q\}+ s(\beta,\Sigma_k)+ \eta, \label{eq:upper_bound_on_Jk}
\end{align}
where the first inequality holds by optimality.
From the definitions of $P_{\xhat}$, ${\bf S}_{\Phi}$ and $Q_b$, we have $P_{\xhat} \succeq Q$. This implies that, if a scalar $\sigma$ exists such that \eqref{eq:condition on sigma} is satisfied, then it must be greater than or equal to $1$. Thus the third inequality in \eqref{eq:upper_bound_on_Jk} follows. A possible choice for $\sigma$ is the largest eigenvalue of $P_{\xhat}Q^{-1}$ if $Q\succ 0$.

We are now able to identify a parametric bound on the averaged undiscounted closed loop cost accumulated over an infinite horizon
in Theorem \ref{thm:parametric bound}.
\begin{thm}\label{thm:parametric bound}
Provided that optimisation \eqref{eq:opt2} is feasible at time instant $k=0$ and a scalar $\sigma$ exists such that \eqref{eq:condition on sigma} is satisfied,
if the discount factor $\beta \in (0,1)$ in optimisation \eqref{eq:opt2}  satisfies
\begin{equation}
\sigma < \frac{1}{1-\beta} \label{eq:condition on beta}
\end{equation} and 
the random sequence $\{\Sigma_k\}_{k=0}^\infty$ is upper bounded by some matrix $\bar{\Sigma}$ (which may depend on $\Sigma_0$), 
then
\begin{equation}
\lim_{T\to\infty}\frac{1}{T}\sum_{k=0}^{T-1} \EE\bigl\{ \| x_{k} \|_Q^2  + \| u_{k}\|_R^2 \bigr\}  \leq  \frac{\sup\limits_{0 \preceq \Sigma \preceq \bar{\Sigma} } s(\beta,\Sigma)+\eta }{\frac{1}{1-\beta}-\sigma}. \label{eq:parametric bound}
\end{equation}
\end{thm}
\vspace{-\baselineskip}
\bpf
From \eqref{eq:equality_cl_cost}, it follows that
\begin{equation}
\beta \EE_k \{J_{k+1}\} \leq J_k-\EE_k\{\norm{x_k}^2_Q+\norm{u_k}^2_R\} \nonumber.
\end{equation}
Multiplying both sides of this inequality by $\beta^{-1}$ and then subtracting
$J_k$ from both sides, we obtain
\begin{equation}
 \EE_k \{ J_{k+1} \} - J_k \!\leq\! (\tfrac{1}{\beta}-1)J_k -\tfrac{1}{\beta} \EE_k\{\norm{x_k}^2_Q+\norm{u_k}^2_R\} \label{eq:ineq_cl_cost}.
\end{equation}
From \eqref{eq:upper_bound_on_Jk} and \eqref{eq:ineq_cl_cost}, it can be concluded that
\begin{align*}
&\EE_k \{ J_{k+1} \} -J_k \leq \bigl((\beta^{-1}-1)\sigma-\beta^{-1}\bigr) \EE_k\{\norm{x_k}^2_Q\} \\
&~ -\beta^{-1} \EE_k\{\norm{u_k}^2_R\} +(\beta^{-1}-1) (s(\beta,\Sigma_k)+\eta ).
\end{align*}
Since $(\beta^{-1}-1)\sigma >0$, we then have
\begin{align}
&( \beta^{-1}\!-\!(\beta^{-1}\!-\!1)\sigma )\EE_k\{ \norm{x_k}^2_Q\!+\!\norm{u_k}^2_R \} \leq   J_k \!-\!\EE_k \{ J_{k+1} \}  \nonumber
 \\
&\hspace{3cm}+ (\beta^{-1}\!-\!1) \bigl(s(\beta,\Sigma_k)+\eta \bigr). \label{eq:new monotonicity condition on Jk}
\end{align}
Furthermore, \eqref{eq:condition on beta} implies that $\beta^{-1}-(\beta^{-1}-1)\sigma>0$.
Summing both sides of \eqref{eq:new monotonicity condition on Jk} over $k=0,1,\ldots,T-1$, dividing by $T$ and considering the limit as $T\!\to\!\infty$, we get
\begin{multline*}
\lim_{T\to\infty}\!\frac{1}{T}\!\sum_{k=0}^{T-1} \!\EE \bigl\{ \| x_{k} \|_Q^2  \!+\! \| u_{k}\|_R^2 \bigr\} \!\leq\!  
\frac{1}{\frac{1}{1-\beta}\!-\!\sigma}\Bigl( 
\eta\! \\
+\!  \lim_{T\to\infty}\!\dfrac{1}{T}\!\sum_{k=0}^{T-1}\EE\{s(\beta,\Sigma_k)\}
\Bigr).
\end{multline*}
Also, from $\Sigma_k \preceq \bar{\Sigma}$ $\forall k \geq 0$ it follows that
\[\lim_{T\to\infty}\frac{1}{T}\sum_{k=0}^{T-1}\EE\{s(\beta,\Sigma_k)\} \leq \sup_{0 \preceq \Sigma \preceq \bar{\Sigma}} s(\beta,\Sigma).\]
By the definition of function $s(\beta,\Sigma)$, this supremum is finite,
which completes the proof.
\epf
\begin{rem}
Theorem \ref{thm:parametric bound} provides a sufficient condition for~\eqref{eq:parametric bound}, which requires $\Sigma_k \preceq \bar{\Sigma}$
$\forall k\geq 0$.
Although not ensured by mean-square stability alone, this requirement is reasonable in real world applications. It can be ensured,
for example, if a common quadratic Lyapunov equation \citep{Lin:09} exists for subsystems of $\xi_{i+1} = A(I - \gamma_iMC) \xi_i$, that is, if there exists a matrix $P=P^\top \succ 0$ such that
\begin{equation}
\begin{aligned}
&P-A^\top PA \succ 0 ,\\
&P-(A-AMC)^\top P(A-AMC) \succ 0
\end{aligned} \label{eq:common quadratic Lyap}
\end{equation}
hold simultaneously. 
\end{rem}

\begin{rem} \label{remark:linear bound on function s}
We can relax the condition on $\Sigma_k$ in Theorem \ref{thm:parametric bound} to mean-square stability of $\Psi(\gamma_k)$ (which is assumed in Assumption \ref{assumption:stabilising_gains}), if there exist some positive constant $\alpha$ and symmetric positive definite matrix $P$ such that
\begin{equation} \label{eq:bound function s}
s(\beta,\Sigma_k) \leq \alpha \tr(\Sigma_k P) \quad \forall \Sigma_k \succeq 0.
\end{equation}
Condition \eqref{eq:bound function s} holds, for example, if a restricted version of the predicted control policy (\ref{eq:predicted_control_law}a) is used, where $L_{i,j|k}$ in (\ref{eq:predicted_control_law}b) is $0$ for all $i=0,\ldots, N-1$, so that ${\bf L}_k$ is not a decision variable in online MPC optimisation problems.
\end{rem}

\subsection{Closed loop cost bound via asymptotic limits} \label{sec:cl_bound_via_limits}
In this section, we allow the discount factors $\beta_1$ and $\beta_2$ in the objective and the constraint in problem \eqref{eq:orig_opt_problem} to differ. The corresponding online optimisation problem defining $\theta_k^\ast$ then becomes
\begin{equation} \label{eq:opt with different betas}
\begin{aligned}
 \min_{\theta_k,P_{\beta_1\!,k},P_{\beta_2\!,k}}
& \trloc (\Q_{\beta_1}{\bf X}_k) + \trloc(\R_{\beta_1}{\bf U}_k) + \trloc(Z_1 P_{\beta_1,k})
\\
\text{s.t.} \quad
& \begin{aligned}[t]
&
\trloc({\bf H}_{\beta_2} {\bf X}_k)
+
\trloc( Z_2  P_{\beta_2,k} )
\leq \mu_{k},
\\
& P_{\beta_1,k} \succeq \beta_1 \EE\bigl\{ \tilde{\Psi}(\gamma) P_{\beta_1,k} \tilde{\Psi}^\top(\gamma) \bigr\} \!+\! \beta_1^N X_{N|k} \\
& ~\quad+ \frac{\beta_1^{N+1}}{1-\beta_1} \EE\bigl\{ \tilde{D}(\gamma) \Bigl[\begin{smallmatrix} \!\Sigma_v\! & \\ & \!\Sigma_w\!\end{smallmatrix} \Bigr] \tilde{D}^{\top}\!(\gamma) \bigr\} , \\
& P_{\beta_2,k} \succeq \beta_2 \EE\bigl\{ \tilde{\Psi}(\gamma) P_{\beta_2,k} \tilde{\Psi}^\top(\gamma) \bigr\} \!+\! \beta_2^N X_{N|k} \\
& ~\quad+ \frac{\beta_2^{N+1}}{1-\beta_2} \EE\bigl\{ \tilde{D}(\gamma) \Bigl[\begin{smallmatrix} \!\Sigma_v\! & \\ & \!\Sigma_w\!\end{smallmatrix} \Bigr] \tilde{D}^{\top}\!(\gamma) \bigr\} ,
\end{aligned}
\end{aligned}
\end{equation}
where $\Q_{\beta_1}$, $\R_{\beta_1}$, ${\bf H}_{\beta_2}$ are constant matrices similarly constructed to $\Q_{\beta}$, $\R_{\beta}$, ${\bf H}_{\beta}$.
To analyse the averaged undiscounted closed loop cost accumulated over an infinite horizon, we keep the discount factor $\beta_2$ fixed, while taking the
left-hand limit of the discount factor in the cost at $\beta_1=1^-$ and then solving the optimisation \eqref{eq:opt with different betas}. Considering the limit at $\beta_1=1^-$ implies that $\beta_1<1$ and hence \eqref{eq:opt with different betas} remains solvable.

We next give a bound on the closed loop cost.
\begin{thm}\label{theorem:bound with beta to 1}
Provided the optimisation problem \eqref{eq:opt with different betas} is feasible at time instant $k=0$, if the discount factor $\beta_1\in (0,1)$ in \eqref{eq:opt with different betas} is arbitrarily close to $1$, then the trajectories of \eqref{eq:system} under the control law of Section \ref{sec:control_implementation} satisfy
\begin{equation}
\lim_{T\to\infty}\frac{1}{T}\sum_{k=0}^{T-1} \EE \bigl\{ \| x_{k} \|_Q^2  + \| u_{k}\|_R^2 \bigr\} \leq \trloc(Z_1 \bar{X}). \label{eq:cl_undis_cost_bound}
\end{equation}
\end{thm}
\vspace{-\baselineskip}
\bpf
We denote the minimiser to \eqref{eq:opt with different betas} with $\beta_1$ being arbitrarily close to $1$ from the left as
$(\hat{\theta}_k^\ast,\hat{P}_k^\ast, \tilde{P}_k^\ast)$,
where $\hat{P}_k^\ast$ and $\tilde{P}_k^\ast$ are the terminal matrices in the predicted cost and the discounted sum constraint, respectively,
and $\hat{{\bf X}}_k^\ast$, $\hat{{\bf U}}_k^\ast$, $\hat{X}_{i|k}^\ast$ as the quantity of ${\bf X}_k$, ${\bf U}_k$, $X_{i|k}$ that correspond to the minimiser $\hat{\theta}^\ast_k$, respectively.
Let $\hat{P}_k^{(T)} \defeq \sum_{i=N}^{N+T-1} \hat{X}^\ast_{i|k}$. Then we have
\begin{equation}
\hat{P}_k^{(T)} = \sum_{i=N}^{N+T-1} ( \hat{X}^\ast_{i|k} - \bar{X}) + T \bar{X}, \label{eq:sequence P_L}
\end{equation}
where $\bar{X}$, the steady state solution to \eqref{eq:recursion X_i}, exists and is unique by Lemma \ref{lemma:MSS_tPsi}.
Dividing both sides of \eqref{eq:sequence P_L} by $T$ and taking the limit at $T=\infty$, we have
\begin{align}
&\lim_{T \to \infty} \frac{1}{T} \hat{P}_k^{(T)} \!=\lim_{T \to \infty} \frac{1}{T} \!\sum_{i=N}^{N+T-1} ( \hat{X}^\ast_{i|k} \!-\! \bar{X}) \!+\! \bar{X} \!=\! \bar{X} \label{eq:limit operation 1}\\
&= \lim_{T \to \infty} \frac{1}{T} \!\sum_{i=N}^{N+T-1} \!\! \lim_{\beta_1 \to 1^-\!} \beta_1^i \hat{X}^\ast_{i|k} \!=\! \lim_{T \to \infty} \frac{1}{T} \hat{P}_k^\ast. \label{eq:limit operation 2}
 \end{align}
The second equality in \eqref{eq:limit operation 1} follows from the observations that $\hat{X}_{i|k}^\ast$ converges to $\bar{X}$ $\forall k\geq 0$ as $i \to \infty$ and $\sum_{i=N}^{\infty}(\hat{X}_{i|k}^\ast\!-\!\bar{X})$ is finite. We next show the second equality in \eqref{eq:limit operation 2} holds.
Whenever $(\hat{\theta}_k^\ast,\hat{P}_k^\ast)$ satisfies its corresponding LMI constraint in problem \eqref{eq:opt with different betas} with equality, $\hat{P}_k^\ast$ satisfies \eqref{eq:def_Pk}. Also, since we solve problem \eqref{eq:opt with different betas} after taking the limit at $\beta_1\!=\!1^-$, each term of that infinite sum on the RHS of \eqref{eq:def_Pk} is evaluated at $\beta_1\!=\!1^-$.

Taking the limit at $\beta_1=1^-$, we obtain a slightly different version of \eqref{eq:equality_cl_cost} as
\begin{multline*}
 \EE_k \{ J(\hat\theta^\ast_{k+1},\xhat_{k+1},\Sigma_{k+1})\}
\\
\leq J(\hat\theta^\ast_k,\xhat_k,\Sigma_k) - \EE_k\{\|x_{k}\|_Q^2 + \|u_{k}\|_R^2\} ,
\end{multline*}
and summing both sides of this inequality over $k=0,\ldots, T-1$ yields that
\begin{equation}
\sum_{k=0}^{T-1}\! \EE \bigl\{ \| x_{k} \|_Q^2  + \| u_{k}\|_R^2 \bigr\} \!\!\leq\! \!J(\hat\theta_0^\ast,\xhat_0,\!\Sigma_0)- \EE \{\! J(\hat\theta_T^\ast,\xhat_T,\!\Sigma_T) \}. \label{eq:sum_cl_cost_up_to_T}
\end{equation}
Dividing both sides of \eqref{eq:sum_cl_cost_up_to_T} by $T$ and taking the limit at $T=\infty$, we have
\begin{align*}
&\lim_{T\to\infty}\frac{1}{T} \sum_{k=0}^{T-1} \EE \bigl\{ \| x_{k} \|_Q^2  + \| u_{k}\|_R^2 \bigr\} \\
& \leq \lim_{T\to\infty}\frac{1}{T} \bigl(J(\hat\theta_0^\ast,\xhat_0,\Sigma_0) \!-\! \EE \{ J(\hat\theta_T^\ast,\xhat_T,\Sigma_T) \}\bigr)\\
&\leq\lim_{T\to\infty}\frac{1}{T} \bigl( \trloc \bigl(\Q_{\beta_1} \hat{{\bf X}}^\ast_0\bigr) + \trloc\bigl(\R_{\beta_1} \hat{{\bf U}}^\ast_0\bigr) + \trloc\bigl(Z_1 \hat{P}_0^\ast\bigr) \bigr) \\
&=\lim_{T\to\infty}\frac{1}{T} \trloc(Z_1 \hat{P}_0^\ast)=  \trloc(Z_1 \lim_{T\to\infty}\frac{1}{T} \hat{P}_0^\ast),
\end{align*}
and this, together with \eqref{eq:limit operation 1} and \eqref{eq:limit operation 2}, implies \eqref{eq:cl_undis_cost_bound}.
\epf

Theorem \ref{theorem:bound with beta to 1} provides some insights into the role the discount factors play. With discount factors $\beta_1, \beta_2\in(0,1)$, we not only ensure the objective and constraint in problem \eqref{eq:orig_opt_problem} take finite values and thus are well-defined despite possibly unbounded disturbances and measurement noise, but we also obtain a trade-off between transient and steady state behaviours. When $\beta_1$ takes values far from~$1$, greater emphasis is put on the cost accumulated over a short horizon near to the initial time, and the costs corresponding to later times are more heavily discounted. On the other hand, as $\beta_1$ approaches $1$, the closed loop cost is dominated by steady state behaviours.

\section{Robustness analysis of the MPC controller} \label{sec:robustness analysis}
This section provides an analysis of the robustness of the MPC controller \eqref{eq:mpc controller} with respect to uncertainties in the arrival probability, $\PP\{\gamma_k=1\}$, of sensor measurements. 
Supposing that only a nominal value, $\lambda_n$, of the arrival probability of sensor measurements is given, while its actual value, $\lambda_a:= \PP\{\gamma_k=1\}$, is time-invariant but unknown and possibly different from $\lambda_n$, we analyse the impact of the error $\Delta \lambda:=\lambda_a -\lambda_n$ on closed loop properties \eqref{eq:cl_constraint}-\eqref{eq:cl_cost}, investigating how one can choose an appropriate value for $\mu_0$ (different from that specified in \eqref{eq:mu_def}) so as to increase the robustness margin. 
To facilitate the following analysis, we define an auxiliary Bernoulli process $\{\widetilde{\gamma}_k\}_{k=0}^\infty$ with $\PP\{\widetilde\gamma_k=1\}=\lambda_n$ and $\PP\{\widetilde\gamma_k=0\}=1-\lambda_n$, and we write $\widetilde\gamma_k \sim \mathcal{B}(1,\lambda_n)$ where $\mathcal{B}(1,\lambda_n)$ denotes a Bernoulli distribution with a success rate $\lambda_n$.

The nominal value of the arrival probability of sensor measurements is used to formulate optimisation \eqref{eq:opt} defining $\theta_k^\ast$ as problem \eqref{eq:opt2} and equivalently as problem \eqref{eq:opt3}. More specifically, $\lambda_n$ is used in computing ${\bs \Omega}_k$ in \eqref{eq:cov_mat} and the terminal matrix $P_k$ in \eqref{eq:term_cost_sdp}. However, the controller implementation and the updates of $\xhat_k$, $\Sigma_k$ and $\mu_k$ depend on realisations of $\gamma_k$. Hence the random variable $\gamma_k$ in \eqref{eq:mpc controller}, \eqref{eq:cl loop priori estimate update}, \eqref{eq:cl_varx} and \eqref{eq:c_tail} assumes values in $\{0,1\}$ according to the probability distribution $\mathcal{B}(1,\lambda_a)$.
\begin{assumption} \label{assummption: nominal and actual greater than threshold}
(i) $(A,\Sigma_w^{\frac{1}{2}})$ is controllable;
(ii) The nominal arrival probability $\lambda_n$ 
is greater than the minimum value \citep[e.g.][]{sinopoli:04} such that there exists a gain matrix $M_1$, ensuring the system
$\xi_{k+1} = (A - \widetilde\gamma_k AM_1C) \xi_k$ is MSS; 
(iii) $\lvert \Delta \lambda\rvert$ is small so that the actual arrival probability $\lambda_a ~(=\Delta\lambda+\lambda_n)$ is sufficiently large such that there exists a gain matrix $M_2$, ensuring the system
$\xi_{k+1} = (A - \gamma_k AM_2C) \xi_k$ is MSS.
\end{assumption}
By exploiting Theorem 5 and Lemma 1 in \cite{sinopoli:04} (which hold by Assumptions \ref{assumption:ctrb_and_obsv} and \ref{assummption: nominal and actual greater than threshold}(i)), we can show that: if there exists a gain matrix $M'$ such that the system
$\xi_{k+1} = (A - \gamma'_k AM'C) \xi_k$ is MSS, where $\{\gamma'_k\}_{k=0}^\infty$ is i.i.d. with $\gamma'_k \sim \mathcal{B}(1,\lambda')$, then, with the same gain matrix, the system $\xi_{k+1} = (A - \gamma''_k AM'C) \xi_k$ is also MSS for any Bernoulli process $\{\gamma''_k\}_{k=0}^\infty$ that is i.i.d. with $\PP\{\gamma''_k=1\} \geq \lambda'$ and $\PP\{\gamma''_k=0\} \leq 1-\lambda'$. Hence, by Assumption \ref{assummption: nominal and actual greater than threshold}, there necessarily exists a gain matrix $M$ such that the systems $\xi_{k+1} = (A - \widetilde\gamma_k AMC) \xi_k$ and $\xi_{k+1} = (A - \gamma_k AMC) \xi_k$ are MSS. 

We next define a simplistic parameterisation of the predicted control sequence as
\begin{subequations} \label{eq:simplistic predicted control parameterisation}
\begin{align}
u_{i|k}&=K \xhat_{i|k}+c_{i|k}, & &i=0, \ldots, N-1,\\
u_{i|k}&=K \xhat_{i|k},         & &i=N, N+1, \ldots.
\end{align}
\end{subequations}
Note that \eqref{eq:simplistic predicted control parameterisation} is a special case and a restricted version of (\ref{eq:predicted_control_law}a,b) with 
$L_{i,j|k}=0$ $\forall i \geq 0$. 
Nevertheless, using \eqref{eq:simplistic predicted control parameterisation} to formulate optimisation \eqref{eq:opt} defining $\theta^\ast_k$, we retain the results on the closed loop system in Sections \ref{sec:properties} and \ref{sec:strengthened stability}.

The main result of this section is given below. 

\begin{thm}\label{theorem:robustness with respect to lambda}
Under Assumption \ref{assummption: nominal and actual greater than threshold}, if parameterisation \eqref{eq:simplistic predicted control parameterisation} is used to formulate optimisation \eqref{eq:opt} defining $\theta^\ast_k$, then, in the closed loop operation of the MPC controller \eqref{eq:mpc controller}, a small uncertainty $\Delta \lambda$ in the arrival probability of sensor measurements results in finite changes to the constraint bound (RHS of (\ref{eq:cl_constraint})) and the cost bound (RHS of (\ref{eq:cl_cost})) that depend linearly on $\Delta\lambda$.
\end{thm}
\vspace{-\baselineskip}
\bpf
We begin the proof with recalling the basis upon which bounds \eqref{eq:cl_constraint} and \eqref{eq:cl_cost} are derived when we have perfect knowledge about the probability distribution of $\gamma_k$, namely $\lambda_n=\lambda_a$. We next introduce intermediate variables $\mu_k^n$ to reconstruct this basis when $\Delta \lambda \neq 0$ and $\lambda_n \neq \lambda_a$. Then the proof is completed by quantifying the resulting changes in \eqref{eq:cl_constraint} and \eqref{eq:cl_cost} from $\Delta \lambda$ as discounted expectations of $\mu_k-\mu_k^n$ accumulated over the infinite horizon.
The proof of Theorem \ref{thm:cl_constraint} is based on the observation that,
at time instant $k$,
the distributions of the state and control sequences $\{x_{i|k+1}\}_{i=0}^\infty$ and $\{u_{i|k+1}\}_{i=0}^\infty$ are identical to the distributions of $\{x_{i+1|k}\}_{i=0}^\infty$ and $\{u_{i+1|k}\}_{i=0}^\infty$.
However, this property no longer holds if $\lambda_n \neq \lambda_a$ since $\gamma_{i|k} \sim \mathcal{B}(1,\lambda_n)$ is assumed, whereas $\gamma_k \sim \mathcal{B}(1,\lambda_a)$.
Given the definition of $\mu_k$ in \eqref{eq:mu_def}, we define a function $h$ such that $\mu_{k+1}=h(\theta^\tail_{k+1},\xhat_{k+1},\Sigma_{k+1})$. We then define an auxiliary variable $\mu_{k+1}^n:= h(\theta^n_{k+1},\xhat^n_{k+1},\Sigma^n_{k+1})$ $\forall k\geq 0$ where $\theta^n_{k+1}:=\{{\bf c}_{k+1}^n, {\bf L}_{k+1}^n\}$ with
\begin{equation}
{\bf c}_{k+1}^n:=
\mathcal{T} {\bf c}_k^\ast
+
\widetilde{\bf L}^\ast_{0|k} \widetilde\gamma_k (y_k - C\xhat_k ) , \label{eq:nominal c}
\end{equation}
and
${\bf L}_{k+1}^n:={\bf L}_{k+1}^\tail$. Here $\xhat^n_{k+1}$ and $\Sigma^n_{k+1}$ are defined as
\begin{align}
&\xhat^n_{k+1} \!:=\! \Phi \xhat_k \!\!+\!\! B c^\ast_{0|k} \!\!+\!\! (AM\!+\!BL^\ast_{0,0|k})\widetilde\gamma_k(y_k \!-\! C\xhat_k),  \label{eq:nomial xhat}
\\
&\Sigma^n_{k+1}\!:=\! (A-\widetilde\gamma_k AMC) \Sigma_{k} (A-\widetilde\gamma_k AMC)^\top \nonumber
\\
&  \qquad\qquad\qquad   + \widetilde\gamma_{k} AM\Sigma_v M^\top A^\top + D \Sigma_w D^\top . \label{eq:nomial sigma_k}
\end{align}
Note that equations \eqref{eq:nominal c}-\eqref{eq:nomial sigma_k} are defined similarly to their counterparts \eqref{eq:c_tail}, \eqref{eq:cl loop priori estimate update} and \eqref{eq:cl_varx} respectively. 
By the definition of function $h$, 
\begin{equation}  \label{eq:made up mu}
\mu_{k+1}^n=\sum_{i=0}^\infty \beta^i \EE_{k+1} \{\|H x_{i|k+1}\|^2 \},
\end{equation}
where the predicted state and corresponding control sequences in \eqref{eq:made up mu} are defined by $\theta^n_{k+1}=\{{\bf c}_{k+1}^n, {\bf L}_{k+1}^n\}$ with initial estimate $\EE_{k+1}\{x_{0|k+1}\}=\xhat^n_{k+1}$ and initial estimation error covariance $\EE_{k+1}\{(x_{0|k+1}-\xhat^n_{k+1})(x_{0|k+1}-\xhat^n_{k+1})^\top\}=\Sigma_{k+1}^n$.
Since the probability distribution of $\widetilde\gamma_k$ is consistent with the imperfect knowledge of the probability distribution of $\gamma_{i|k}$, we recover the equivalence that,
at time instant $k$,
the distributions of the state and control sequences $\{x_{i|k+1}\}_{i=0}^\infty$ and $\{u_{i|k+1}\}_{i=0}^\infty$ that are associated with $\theta^n_{k+1}$, $\xhat_{k+1}^n$ and $\Sigma^n_{k+1}$ are identical to the distributions of $\{x_{i+1|k}\}_{i=0}^\infty$ and $\{u_{i+1|k}\}_{i=0}^\infty$. Therefore,
\[
\EE_k\{\mu_{k+1}^n\} =\sum_{i=0}^\infty \beta^i \EE_{k} \{\|H x_{i+1|k}\|^2\},
\]
and it follows that
\begin{align}
&\beta\EE_k\{\mu_{k+1}\}=\beta\EE_k\{\mu_{k+1}-\mu_{k+1}^n\} + \beta\EE_k\{\mu_{k+1}^n\} \nonumber 
\\
&\negmedspace{}=\beta\EE_k\{\mu_{k+1}-\mu_{k+1}^n\} + \sum_{i=0}^\infty \beta^{i} \EE_{k} \{\|H x_{i|k}\|^2\} \nonumber
\\
& \hspace{5.8cm} -\EE_{k} \{\|H x_{0|k}\|^2\} \nonumber
\\
&\negmedspace{}\leq \beta\EE_k\{\mu_{k+1}-\mu_{k+1}^n\} +\mu_k- \EE_k\{\|Hx_{k}\|^2\}. \label{eq:imperfect ineq to cl_constraint}
\end{align}
Summing both sides of \eqref{eq:imperfect ineq to cl_constraint} over $k=0, 1, \ldots$, we have
\begin{equation}
\sum_{k=0}^\infty \beta^k \EE\{\norm{H x_k}^2\} \leq \mu_0 + \sum_{k=0}^\infty \beta^{k+1} \EE\{\mu_{k+1}-\mu_{k+1}^n \} \nonumber.
\end{equation}
To quantify $\sum_{k=0}^\infty \beta^{k+1} \EE\{\mu_{k+1}-\mu_{k+1}^n \}$, we next compute $\EE_k\{\mu_{k+1}-\mu_{k+1}^n\}$. To streamline the presentation of this proof, the detailed algebraic calculation is omitted while noting that $\EE_k\{\mu_{k+1}\}$ can be computed using $X_{N|k+1}^\tail$ and ${\bf X}_{k+1}^\tail$ defined in \eqref{eq:feasible Pk and X_k} and $\EE_k\{\mu_{k+1}^n\}$ can be computed using $X_{N|k+1}$ and ${\bf X}_{k+1}$ that depend on $\theta^n_{k+1},~\xhat^n_{k+1}$ and $\Sigma^n_{k+1}$. We obtain $\EE_k\{\mu_{k+1}-\mu_{k+1}^n\}$ as
\begin{equation}
\EE_k\{\mu_{k+1}\}-\EE_k\{\mu_{k+1}^n\}=\Delta\lambda\,\mathcal{L}_{k+1}(\Sigma_k), \nonumber
\end{equation}
where
\begin{scriptsize}
\[
\begin{aligned}
&\mathcal{L}_{k+1}(\Sigma_k)
\\
&\negmedspace{}:=\tr \Bigl( H_b  ({\bf S}_\Phi (AM\!+\!BL^\ast_{0,0|k}) +{\bf T}_{(\Phi,B)} \widetilde{\bf L}^\ast_{0|k} ) (C\Sigma_kC^\top \!+\Sigma_v) 
\\
&\negmedspace{}\cdot\,({\bf S}_\Phi (AM+BL^\ast_{0,0|k}) +{\bf T}_{(\Phi,B)} \widetilde{\bf L}^\ast_{0|k} )^\top  \Bigr) + \tr \Bigl( \bigl\|[I~ {\bs \Pi}_{k+1}^\tail]\bigr\|^2_{H_b}
\\
&\negmedspace{}\cdot\,\sum_{j}
\Bigl[
\begin{smallmatrix}
F({\bs \Gamma}^{(j)}) \\
G({\bs \Gamma}^{(j)})
\end{smallmatrix}
\Bigr]
\Delta_k
\Bigl[
\begin{smallmatrix}
F({\bs \Gamma}^{(j)}) \\
G({\bs \Gamma}^{(j)})
\end{smallmatrix}
\Bigr]^\top
\PP\{{\bs \Gamma}_{k+1}={\bs \Gamma}^{(j)}\} \Bigr)
\\
&\negmedspace{}+\!\beta^N \!\tr \Bigl( \tilde{Z}_{2,22}  ({S}^N_\Phi (AM\!\!+\!\!BL^\ast_{0,0|k}) \!+\!{T}^N_{(\Phi,B)} \widetilde{\bf L}^\ast_{0|k} ) (C\Sigma_kC^\top \!\!+\!\Sigma_v) 
\\
&\negmedspace{}\cdot\,({S}^N_\Phi (AM\!\!+\!\!BL^\ast_{0,0|k}) \!+\!{T}^N_{(\Phi,B)} \widetilde{\bf L}^\ast_{0|k} )^\top  \Bigr) + \beta^N \tr \Bigl( \Bigl\|\! \Bigl[\!\begin{smallmatrix}I \!&  \\  & \Pi_{N|k+1}^\tail\end{smallmatrix}\!\Bigr] \!\Bigr\|^2_{\tilde{Z}_2}  
\\
&\negmedspace{}\cdot\,\sum_{j}
\Bigl[
\begin{smallmatrix}
F_N({\bs \Gamma}^{(j)}) \\
G({\bs \Gamma}^{(j)})
\end{smallmatrix}
\Bigr]
\Delta_k
\Bigl[
\begin{smallmatrix}
F_N({\bs \Gamma}^{(j)}) \\
G({\bs \Gamma}^{(j)})
\end{smallmatrix}
\Bigr]^\top
\PP\{{\bs \Gamma}_{k+1}={\bs \Gamma}^{(j)}\} \Bigr),
\end{aligned}
\]
\end{scriptsize}%
and $\Delta_k:=
\diag \{
(A-AMC)\Sigma_k(A-AMC)^\top \!+AM \Sigma_v M^\top A^\top \!-A\Sigma_k A^\top, 0, 0
\}.$ Therefore, 
we have
\begin{equation}\label{eq:perturbed cl constraint bound}
\sum_{k=0}^{\infty} \beta^k \EE\{\|H x_{k}\|^2\} \leq \mu_0+ \Delta \lambda\sum_{k=0}^\infty \beta^{k+1}\EE\{\mathcal{L}_{k+1}(\Sigma_k)\}.
\end{equation}
It is clear that if $\lambda_a=\lambda_n$, \eqref{eq:perturbed cl constraint bound} is identical to \eqref{eq:cl_constraint}. If parameterisation \eqref{eq:simplistic predicted control parameterisation} is used to formulate optimisation \eqref{eq:opt} defining $\theta^\ast_k$, then $L_{0,0|k}^\ast$, $\widetilde{\bf L}_{0|k}^\ast$, ${\bs \Pi}^\tail_{k+1}$ and $\Pi^\tail_{N|k+1}$ are  constant matrices and $\mathcal{L}_{k+1}$ becomes a linear time-invariant function of $\Sigma_k$. Moreover, Assumption \ref{assummption: nominal and actual greater than threshold} implies that there exists some matrix $\bar\Sigma$ such that $\EE\{\Sigma_k\} \preceq \bar\Sigma$ $\forall k \geq 0$. Therefore, the resulting change in bound \eqref{eq:cl_constraint} from $\Delta \lambda$ can be upper bounded by some finite constant depending linearly on $\Delta\lambda$ and $\bar\Sigma$. The analysis of the cost function to show that $\Delta \lambda\neq 0$ causes a finite change in the bound \eqref{eq:cl_cost} is similar and therefore is omitted.
\epf


Instead of setting $\mu_0:=\epsilon$ as in \eqref{eq:mu_def}, it is recommended that $\mu_0$ is chosen conservatively if $\PP\{\gamma_k=1\}$ is not known exactly. In particular, we require
\[
\Delta \lambda\sum_{k=0}^\infty \beta^{k+1}\EE\{\mathcal{L}_{k+1}(\Sigma_k)\} \leq \epsilon -\mu_0
\]
so that the bound \eqref{eq:cl_constraint} remains valid, which suggests that $\mu_0$ should be chosen as the minimal value such that problem \eqref{eq:opt2} is initially feasible.
Then the gap between $\epsilon$ and $\mu_0$ provides a robustness margin to allow for uncertainty in the arrival probability of sensor measurements.


\section{Numerical examples}\label{sec:example}
In this section, we run three sets of simulations: (A)~demonstrates that the closed loop system satisfies the bounds in \eqref{eq:cl_constraint} and \eqref{eq:cl_cost}, and compares performance with the unconstrained optimal LQG controller; (B)~shows that the averaged undiscounted closed loop cost accumulated over an infinite horizon is finite under the conditions of Theorem~\ref{thm:parametric bound}; and (C)~demonstrates that the closed loop system satisfies~\eqref{eq:cl_undis_cost_bound}.

\textit{Simulation A }: We consider a system obtained by discretising a linearised continuous time model of a double inverted pendulum with a sample time of 0.01\,s as  in \cite{Davison:90}.   The system matrices are
\begin{gather*}
A=
\left[\begin{smallmatrix}
    1.0005  &  0.01 &  -0.0005  &  0 \\
    0.098  &  1.0005 &  -0.0981  & -0.0005 \\
   -0.0005  &  0 &   1.0015  &  0.01 \\
   -0.0981  & -0.0005 &   0.2942  &  1.0015
\end{smallmatrix}\right], \\
B =
\left[\begin{smallmatrix}
    0.0001 &  -0.0001\\
    0.01 &  -0.02    \\
   -0.0001 &   0.0003\\
   -0.02 &   0.05
\end{smallmatrix}\right],
\quad
C=
\left[\begin{smallmatrix}
     1 &    0 &    0 &    0\\
     0 &    0 &    1 &    0
\end{smallmatrix}\right],
\quad
D=I,
\end{gather*}%
and $\omega_k \sim \mathcal{N}(0,\Sigma_w)$, $v_k \sim \mathcal{N}(0,\Sigma_v)$, $\lambda=0.6$. Here $\Sigma_w=\diag\{0.5, 0.2 , 0.9, 0.3\}$ and $\Sigma_v=1.1 I$. Initial conditions are given by
\begin{equation*}
x_0=\left[\begin{smallmatrix}  -0.8\\ 0.4\\ 0.55\\ -0.5 \end{smallmatrix}\right], \
\xhat_0= \left[\begin{smallmatrix} 0.1\\ 0.05\\ 0.1\\ 0.05 \end{smallmatrix}\right], \
\Sigma_0=
\left[\begin{smallmatrix}
    0.5 &  -0.5 &  -0.5 &  0.5\\
   -0.5 &   0.5 &  0.5  & -0.5\\
   -0.5 &   0.5 &  0.5  & -0.5\\
    0.5 &  -0.5 & -0.5  &  0.5
\end{smallmatrix}\right].
\end{equation*}%
The constraint of \eqref{eq:orig_opt_problem} is defined by $\beta=0.8$, $\epsilon=2$ and
\[H=
\left[ \begin{smallmatrix}
0  &  0.1  &  0 &  -0.1\\
0.1  &  0  & -0.1 &  0
\end{smallmatrix} \right].
\]
The weighting matrices in the cost function of \eqref{eq:orig_opt_problem} are given by
$Q=\diag \{10, 0.1, 10, 0.1\}$, $R=0.01 I$. We choose a prediction horizon $N=5$, $K$ as the unconstrained LQ-optimal, $K_{LQ}$, with respect to $(A,B,Q,R)$, and $M=\hat\Sigma C^\top (C \hat\Sigma C^\top + \Sigma_v)^{-1}$, where $\hat\Sigma$ is the solution of the algebraic Riccati equation
\[
\hat\Sigma=A \hat\Sigma A^\top +\Sigma_w -\lambda A\hat\Sigma C^\top (C\hat\Sigma C^\top +\Sigma_v)^{-1}C\hat\Sigma A^\top.
\]
For this system, cost and constraint, the solution of \eqref{eq:opt2} yields $J_0=9.0757\times 10^5$.

To verify~\eqref{eq:cl_constraint} and \eqref{eq:cl_cost} experimentally, we
consider the average cost and constraint values over $10^3$ simulations, each of which has a run time of $150$ time steps. For comparison, we run the same number of simulations with the same $\{\omega_k\}$, $\{v_k\}$, $\{\gamma_k\}$ sequences using the unconstrained optimal LQG controller, where $u_k=K_{LQ}\xhat_k$ and the estimator gain is time-varying and given by $M=\Sigma_k C^\top (C \Sigma_k C^\top + \Sigma_v)^{-1}$. Here $\Sigma_k$ evolves as
\[
\Sigma_{k+1}\!\!=\!\!A \Sigma_k A^\top \!+\Sigma_w -\gamma_k A\Sigma_k C^\top \!(C\Sigma_k C^\top \!+\Sigma_v)^{-1}C\Sigma_kA^\top\!.
\]
The results summarised in Table~\ref{tab:simA} agree with the bounds in~\eqref{eq:cl_constraint} and \eqref{eq:cl_cost} and show that, although the LQG controller gives a smaller closed loop cost (as expected), it violates the constraint. Note that $\beta^{150} = 2.9074 \times 10^{-15}$, so a further increase in the simulation run time has negligible effect on the cost and constraint estimates.

\begin{table}[h]
\begin{small}
\begin{tabularx}{0.47\textwidth} {
   >{\raggedright\arraybackslash}X
  | >{\centering\arraybackslash}X
  | >{\centering\arraybackslash}X  }

  & MPC controller & LQG controller
  \\
 \hline
 empirical cost  & $\!2.0545 \!\times\! 10^4<J_0$  & $893.5569$
\\
\hline
empirical constraint  & $1.5175$ $<\epsilon$  &  $3.9272>\epsilon$
\end{tabularx}
\end{small}
\vspace{1pt}
\caption{Average discounted cost and constraint values for Simulation~A}
\label{tab:simA}

\end{table}

\textit{Simulation B}: For this simulation the system model, cost and constraint parameters are the same as for Simulation~A, except that the state matrix is redefined as $A \gets A-1.175I$ and the unconstrained LQ-optimal feedback gain $K_{LQ}$, the steady state Kalman filter gain $M$ are modified accordingly. The modified matrices $A$ and $M$ satisfy \eqref{eq:common quadratic Lyap} and there exists a scalar $\sigma=4.2759$ such that \eqref{eq:condition on sigma} holds and $\sigma<\frac{1}{1-\beta}=5$. To estimate empirically the LHS of \eqref{eq:parametric bound}, we run $\smash{10^2}$ simulations, each of which has a run time of $10^4$ time steps, and we thus obtain an estimate of the average cost value from the average over $10^2$ simulations as $14.5375$. For this example therefore, the LHS of \eqref{eq:parametric bound} is finite as implied by Theorem~\ref{thm:parametric bound}. Further increases in the simulation run time cause negligible changes in this estimate.

\textit{Simulation C}: In this simulation we allow for different discount factors $\beta_1$ and $\beta_2(=\beta=0.8)$ in problem~\eqref{eq:orig_opt_problem} and we set $\epsilon=3.8$. The weighting matrix for control inputs is given by $R=0.001 I$ and the unconstrained LQ-optimal feedback gain, $K_{LQ}$, changes accordingly. All other model and problem parameters are the same as for Simulation~A. We solve the steady state equation of \eqref{eq:recursion X_i} and obtain the RHS of \eqref{eq:cl_undis_cost_bound} as $518.3913$. To estimate empirically the LHS of \eqref{eq:cl_undis_cost_bound} as $\beta_1$ approaches $1$ while keeping $\beta_2$ fixed, we run three tests with $\beta_1$ equal to $0.98$, $0.99$ and $0.999$, respectively. Each test consists of $100$ simulations, each of which has a run time of $10^4$ time steps. Empirical values of the LHS of \eqref{eq:cl_undis_cost_bound} corresponding to different values of $\beta_1$ are summarised in Table~\ref{tab:simC}, and are in agreement with the bound \eqref{eq:cl_undis_cost_bound}. Note also that a further increase of the simulation run time only results in small variations in these estimates.

\begin{table}[h]
\begin{small}
\begin{tabularx}{0.47\textwidth} {
   >{\raggedright\arraybackslash}X
  | >{\raggedright\arraybackslash}X
  | >{\raggedright\arraybackslash}X
  | >{\raggedright\arraybackslash}X |
  | >{\raggedright\arraybackslash}X  }
  & $\!\beta_1\!=\!0.98$ & $\!\beta_1\!=\!0.99$ & $\!\!\!\beta_1\!\!=\!0.999$ & $\trloc(Z_1\bar{X})$
  \\
 \hline
 empirical cost  & 483.3557  & 472.8744  &  468.7458 & 518.3913
\end{tabularx}
\end{small}
\vspace{1pt}
\caption{Average undiscounted cost values for Simulation C}
\label{tab:simC}
\end{table}

\section{Conclusion}\label{sec:conclusion}
This paper proposes an output feedback MPC algorithm for linear discrete time systems with additive disturbances and noisy sensor measurements transmitted over a packet-dropping communication channel.
By parameterising a control policy in terms of affine functions of future observations, we provide a convex formulation of a stochastic quadratic regulation problem subject to a discounted expectation constraint.
Our controller ensures recursive feasibility of the MPC optimisation problem and ensures constraint satisfaction and a discounted cost bound in closed loop operation. We provide a sufficient condition on the discount factor to ensure that the averaged undiscounted closed loop cost is finite. 
%
We consider conditions to ensure closed loop stability and investigate the effects of uncertainties in the probability of successfully receiving a sensor measurement.

\bibliographystyle{apalike}
{\small
\bibliography{IEEEabrv,reference}             
}                                    

\end{document}